\documentclass[journal]{IEEEtran}
\usepackage{graphicx}
\usepackage[table,xcdraw]{xcolor}
\usepackage{amsmath}
\usepackage{amssymb}
\usepackage{booktabs}
\usepackage{multirow}
\usepackage{color}
\usepackage{adjustbox}
\usepackage{xcolor}
\usepackage{cite}
\usepackage{algorithm}
\usepackage[vlined, ruled, algo2e]{algorithm2e}
\usepackage{caption}
\usepackage{subcaption}
\usepackage{csquotes}
\usepackage{dblfloatfix}   % this one is for the image align bottom
\usepackage{float}

\usepackage[inline]{enumitem}
\usepackage{hyperref}
\usepackage{varwidth}% http://ctan.org/pkg/varwidth
\makeatletter
\renewcommand{\Indentp}[1]{%
	\advance\leftskip by #1
	\advance\skiptext by -#1
	\advance\skiprule by #1}%
\renewcommand{\Indp}{\algocf@adjustskipindent\Indentp{\algoskipindent}}
\renewcommand{\Indm}{\algocf@adjustskipindent\Indentp{-\algoskipindent}}
\makeatother

\ifodd 1
\newcommand{\revtwo}[1]{{\color{black} #1}} %revision 2 part
\else

\fi
\begin{document}
% Do not put math or special symbols in the title.
\title{Location-based \revtwo{Activity Behavior} Deviation Detection for Nursing Home using IoT Devices}

\author{Billy~Pik~Lik~Lau,~\IEEEmembership{Member,~IEEE,}
		Zann~Koh,
		Yuren~Zhou,~\IEEEmembership{Member,~IEEE,}
		Benny~Kai~Kiat~Ng, 
        Chau~Yuen,~\IEEEmembership{Fellow,~IEEE}, and
        Mui Lang Low
\thanks{Billy Pik Lik Lau, Zann~Koh, Yuren~Zhou, Benny Kai Kiat Ng, and Chau Yuen are with the Engineering Product Development, Singapore University of Technology and Design, Corresponding E-mail: billy\_lau@mymail.sutd.edu.sg, yuenchau@sutd.edu.sg.
	
Mui Lang Low is with the Peacehaven Nursing Home Day Centre run by The Salvation Army.}
\thanks{Manuscript received January 25, 2023}}

% The paper headers
\markboth{Published in Elsevier Internet of Things, 25 January 2023}%
{Shell \MakeLowercase{\textit{et al.}}: Bare Demo of IEEEtran.cls for IEEE Journals}

% make the title area
\maketitle

\begin{abstract}
	With the advancement of the Internet of Things(IoT) and pervasive computing applications, it provides a better opportunity to understand the behavior of the aging population.
	However, in a nursing home scenario, common sensors and techniques used to track an elderly living alone are not suitable.
	In this paper, we design a location-based tracking system for a four-story nursing home - The Salvation Army, Peacehaven Nursing Home in Singapore.
	The main challenge here is to identify the group activity among the nursing home's residents and to detect if they have any deviated \revtwo{activity behavior}.
	We propose a location-based deviated \revtwo{activity behavior} detection system to detect deviated \revtwo{activity behavior} by leveraging data fusion technique.
	In order to compute the features for data fusion, an adaptive method is applied for extracting the group and individual activity time and generate daily hybrid norm for each of the residents.
	Next, deviated \revtwo{activity behavior} detection is executed by considering the difference between daily norm patterns and daily input data for each resident.
	Lastly, the deviated \revtwo{activity behavior} among the residents are classified using a rule-based classification approach.
	Through the implementation, there are 44.4\% of the residents do not have deviated \revtwo{activity behavior}, while 37\% residents involved in one deviated \revtwo{activity behavior} and 18.6\% residents have two or more deviated \revtwo{activity behaviors}.
\end{abstract}


%%%%%%%%%%%%%%%%%%%%%%%%%%%%%%%%%%%%%%%%%%%%%%%%%%%%%%%%%%%%%%%%%%%%%%%%%%%%%%%%%%%%%%%%%%%%%%%

%%%%%%%%%%%%%%%%%%%%%%%%%%%%%%%%%%%%%%%%%%%%%%%%%%%%%%%%%%%%%%%%%%%%%%%%%%%%%%%%%%%%%%%%%%%%%%%%%%%%%%%%%%
\vspace{-0.14cm}
\section{Introduction}
\label{sec:Introduction}
%%%%%%%%%%%%%%%%%%%%%%%%%%%%%%%%%%%%%%%%%%%%%%%%%%%%%%%%%%%%%%%%%%%%%%%%%%%%%%%%%%%%%%%%%%%%%%%%%%%%%%%%%%
Over the past few years, the advancement of the Internet of Things (IoT) technology has opened up a lot of research potential in the area of tracking and monitoring. 
Among them, a wide variety of projects have been carried out to monitor the behavior of the human being as shown in~\cite{Lau2018Sensor, Mighali2017smart, Dhingra2021Internet, Marakkalage2019Identifying, Wang2016outdoor}.
These technologies made room for implementing convenient applications for enhancing day to day living of urban residents. 

%~\cite{Lau2018Sensor, Mighali2017smart, Marakkalage2019Identifying, Wang2016outdoor}.

With the increase of the world aging population as shown in~\cite{becker2015world} and~\cite{kaneda2006china}, research in monitoring the elderly has gained attention from different research principles.
While these works~\cite{Aran2016Anomaly,Rahman2022iRestroom,Sokullu2020IoT,Kurnianingsih2018Detection} address the support of the elderly as independent beings of the society, and others~\cite{Pierleoni2015High,suzuki2006monitoring} have provided the facility for the nursing home to monitor the daily \revtwo{activity behavior} of the residents.
The former method commonly leverages boundary-less tracking and monitoring techniques as shown in~\cite{Marakkalage2019Identifying,Wang2016outdoor}, which include smartphones and smart wearable devices.
The majority of the latter approaches~\cite{Pierleoni2015High,suzuki2006monitoring} mostly provide tracking in a confined environment, where the accuracy of the boundary-less approach is not ideal.
Our work focuses on the latter approach, where the constraints of monitoring senior citizens are limited to a nursing home. 
Traditionally, it is labor-intensive to take care of the day-to-day life of an elderly resident, and it is not possible to constantly track an individual across 24 hours.
Therefore, using a building-scale human monitoring approach, it can assist the nursing home staff to monitor residents and better understand their \revtwo{activity behavior}.

With this challenge in mind, we design a system to monitor the deviated \revtwo{activity behavior} of nursing home's residents leveraging IoT technology.
We use bluetooth low energy (BLE) technology as backbone for collecting the elderly data due to nature of low energy, and coverage suitable for indoor application compared to WiFi, RFID, ZigBee, etc.
The deviated \revtwo{activity behavior} denotes a nursing home's resident behaves irregularly compared to his/her normal routine of daily life.
This type of detection only can be achieved through fully understanding a resident's life routine.
The main objectives of such a system are to identify the residents' \revtwo{ activity behavior} and determine the irregular \revtwo{activity behaviors}. 
The constraints of monitoring residents' \revtwo{activity behavior} in a nursing home are bounded by building structure, and also their daily activity is influenced by the group activities or community.
Therefore, our aim is to differentiate their activity between private and group activity when computing their deviated \revtwo{activity behavior}.
\revtwo{Another constraint when designing this system is that we do not have ground truth on the data collected, which resulting the accuracy of system output cannot be validated.
Moreover, the identity of the nursing home residents is anonymized to comply with Singapore Personal Data Protection Act~\cite{Commission2012Personal}. 
Thus, unsupervised knowledge extraction is more desired when compared to the supervised knowledge extraction model.}

To address the aforementioned challenges, in this paper, we present a building-scale monitoring system to study 50 residents' \revtwo{activity behavior} in the Peacehaven Nursing Home, Singapore.
Using the building-scale monitoring system, residents' \revtwo{activity behavior} based on their location are investigated using a wearable card tag with a build-in Bluetooth beacon.
Each room is equipped with a receiver to detect the Bluetooth beacon transmitted to perform the resident's location detection. 
Based on the detected location, we study the \revtwo{activity behavior} of residents over 6 months and cluster them based on their common patterns. 
In order to identify the normal \revtwo{activity behavior}, we use a data fusion method to generate the hybrid norm by combining the group and individual norm. 
Using the hybrid norm, deviated \revtwo{activity behavior} can be extracted, which does not follow the normal daily pattern of a resident.
Afterward, we perform empirical analysis on the deviated \revtwo{activity behavior} and classify them.

The key contributions of this paper are as follows:
\begin{itemize}
\item We study the resident's \revtwo{activity behavior} in a nursing home from a location-based implementation of a monitoring system.
\item We propose a data fusion method to identify the daily norm for each nursing home's resident based on two data sources, which are individual and group norm.
\item Based on the daily norm generated, we perform empirical analysis on the deviated \revtwo{activity behavior} and identify the types of them using rules-based classification.
\end{itemize}

Our paper can be detailed as follows: In Section~\ref{sec:literatureStudy}, we study related work about existing methodologies in detecting deviated \revtwo{activity behavior} with their pros and cons. 
Subsequently, in Section~\ref{sec:systemDesign}, the system architecture and data processing model is presented.
Afterward, we describe the group \revtwo{activity behavior} clustering method in Section~\ref{sec:groupDetection}. 
Based on the group detected, we compute the deviated \revtwo{activity behavior} utilizing the hybrid norm and analyze them in Section~\ref{sec:irregularStudy}.
Lastly, we conclude our work in Section~\ref{sec:conclusion}.
\vspace{-0.16cm}

%%%%%%%%%%%%%%%%%%%%%%%%%%%%%%%%%%%%%%%%%%%%%%%%%%%%%%%%%%%%%%%%%%%%%%%%%%%%%%%%%%%%%%%%%%%%%%%%%%%%%%%%%%
\section{Related Work}
\label{sec:literatureStudy}
%%%%%%%%%%%%%%%%%%%%%%%%%%%%%%%%%%%%%%%%%%%%%%%%%%%%%%%%%%%%%%%%%%%%%%%%%%%%%%%%%%%%%%%%%%%%%%%%%%%%%%%%%%
In this section, we discuss some of the related works in the field, which are types of monitoring techniques used to achieve human monitoring and methodologies applied to study deviated \revtwo{activity behavior}. 
\vspace{-0.19cm}

\subsection{Types of Monitoring Techniques}
There are four common types of monitoring techniques in the literature, which are (1) people-driven, (2) event-driven, (3) location-driven, and (4) data-driven.

The people-driven monitoring technique uses humans as the main source of generating information, which normally involves sensors installed in smartphones, watch, bracelets, and other types of wearable.
It is commonly not restricted by location and has a wider coverage of sensing capability.
Examples of smartphone monitoring techniques can be found in~\cite{Marakkalage2019Identifying,Ouchi2013Smartphone,Lau2017Extracting}, where common sensors used are accelerometer, GPS, microphone, etc. 
Examples of other types of wearable devices are belt equipped with sensing unit~\cite{Pierleoni2015High} and bracelets~\cite{Kurnianingsih2018Detection}.
Generally, these types of monitoring techniques are intrusive but able to capture good accuracy data.

The event-driven monitoring technique uses the activity of daily life (ADL) of the users and attempts to understand the \revtwo{activity behavior} of the targeted user.
For instance, Alcala et. al.~\cite{alcala2015detecting} uses the hidden Markov model (HMM) to process ADL and detect the deviated \revtwo{activity behavior} from the daily routine, while Zerkouk et.al.~\cite{zerkouk2019long} use a long term short term memory-based model to identify deviated routine among senior citizen.
A detailed review of the ADL monitoring approaches can be found in~\cite{vermeulen2011predicting}.
The downside of the events-driven monitoring technique is that detailed data is desired and requires a lot of effort as incomplete information will mislead the study outcome.   

When monitoring techniques involve installing multiple sensors at a particular location or building, often it is known as a location-driven method of monitoring people. 
The coverage of the monitoring often involves a building or a particular area, and commonly used sensors include motion sensors~\cite{lotfi2012smart}, vision~\cite{Harrou2017Vision}, RFID sensors~\cite{hsu2010rfid}, infra-red~\cite{Gochoo2018Devicea}, WiFi-passive~\cite{Zhou2020Understanding}. 
However, the inconvenience of this monitoring approach is limited to the area coverage since it is location-bound and only limited study scenarios would benefit from such an approach. 

The data-driven approach uses various information sources and combines them to infer human \revtwo{activity behavior}. 
It can be a mixture of different data sources as described in~\cite{Lau2019survey} such as physical sensors or cyber data sources such as social media.
Examples of data fusion driven approaches can be found in these works~\cite{Ghayvat2018Smart, Suryadevara2012Sensor}, where multiple sensors are fused to study human behavior.
The disadvantages of this approach are due to the complexity of the model and domain knowledge required to select relevant information sources to combine. 
Besides, every data sources require different preprocessing techniques, which can be rather tedious and challenging.
\vspace{-0.19cm}

\subsection{Methodologies in Deviated \revtwo{Activity Behavior} Extraction}
The common methodologies in studying the deviated \revtwo{activity behavior} of senior citizens can be categorized into the following: (1) prediction model, (2) state estimation model, and (3) clustering and exploration model.

The prediction model utilizes statistics to predict the potential \revtwo{activity behavior} of a particular user and if the predicted behavior does not match the predictive outcome, it will be labeled as deviated \revtwo{activity behavior}. 
Recent prediction methods such as long short term memory (LSTM) can be found in~\cite{zerkouk2019long}, which detect the deviated \revtwo{activity behaviors} among the senior citizens using a deep learning approach.
Other types of statistical predictive models also can be found in~\cite{lotfi2012smart,Aran2016Anomaly,Zekri2020Using,Shin2011Detection}.
The predictive model generally requires good quality and a large amount of data as it is not ideal to perform a model with limited or noisy data.

State estimation modeling maps the state behavior of a particular user into a system state, which can be used to model the users' \revtwo{activity behavior} and detect any deviated \revtwo{activity behavior}.
State estimation usually requires human intervention to map the total state of the given system, which involves domain experts to carry out such tasks.
In~\cite{Novak2012Unobtrusive}, Novak et. al. performed the Self Organizing Maps (SOM) and Makrov prediction model onto ADL of a senior citizen to detect their deviated \revtwo{activity behavior}.
Another example of state estimation modeling can be found in~\cite{Ishii2018Method}, where they detect the deviated \revtwo{activity behavior} using HMM.
Other state estimation examples can be found in~\cite{monekosso2009anomalous, singla2010recognizing}.
The drawback of this approach can be rather complex since it does not consider the relationship between activities that happened in parallel.

The clustering and exploration model normally use four steps to generate an insights extraction model, which the user's deviated \revtwo{activity behavior} is studied during the exploration phase. 
It is first proposed in \cite{Cheng2006Multiscale} and commonly used when there is no ground-truth available or no prior knowledge regarding deviated \revtwo{activity behavior}. 
An example of this approach can be found in \cite{Kurnianingsih2018Detection}, where Kurnianingsih et. al. use hybrid $k$-means clustering and isolation forest to detect the deviated vital signals among senior citizen.
In~\cite{hsu2010rfid}, authors also use $k$-means clustering to formulate normal pattern and if any event does not fit into the cluster, it will be labeled as deviated \revtwo{activity behavior}.
Meanwhile, authors~\cite{Marakkalage2019Identifying} studied the \revtwo{activity behavior} of senior citizens using $k$-means clustering approach and decision tree to generate features for analyzing the users' demographic. 
The main drawback of this method is that it requires extensive knowledge in specific domains when analyzing potential deviated \revtwo{activity behavior}, however it works effectively when there is no ground-truth available.
\vspace{-0.19cm}

\begin{figure}[h]
\begin{subfigure}[sample1]{0.4\textwidth}
	\centering \includegraphics[width=0.8\textwidth]{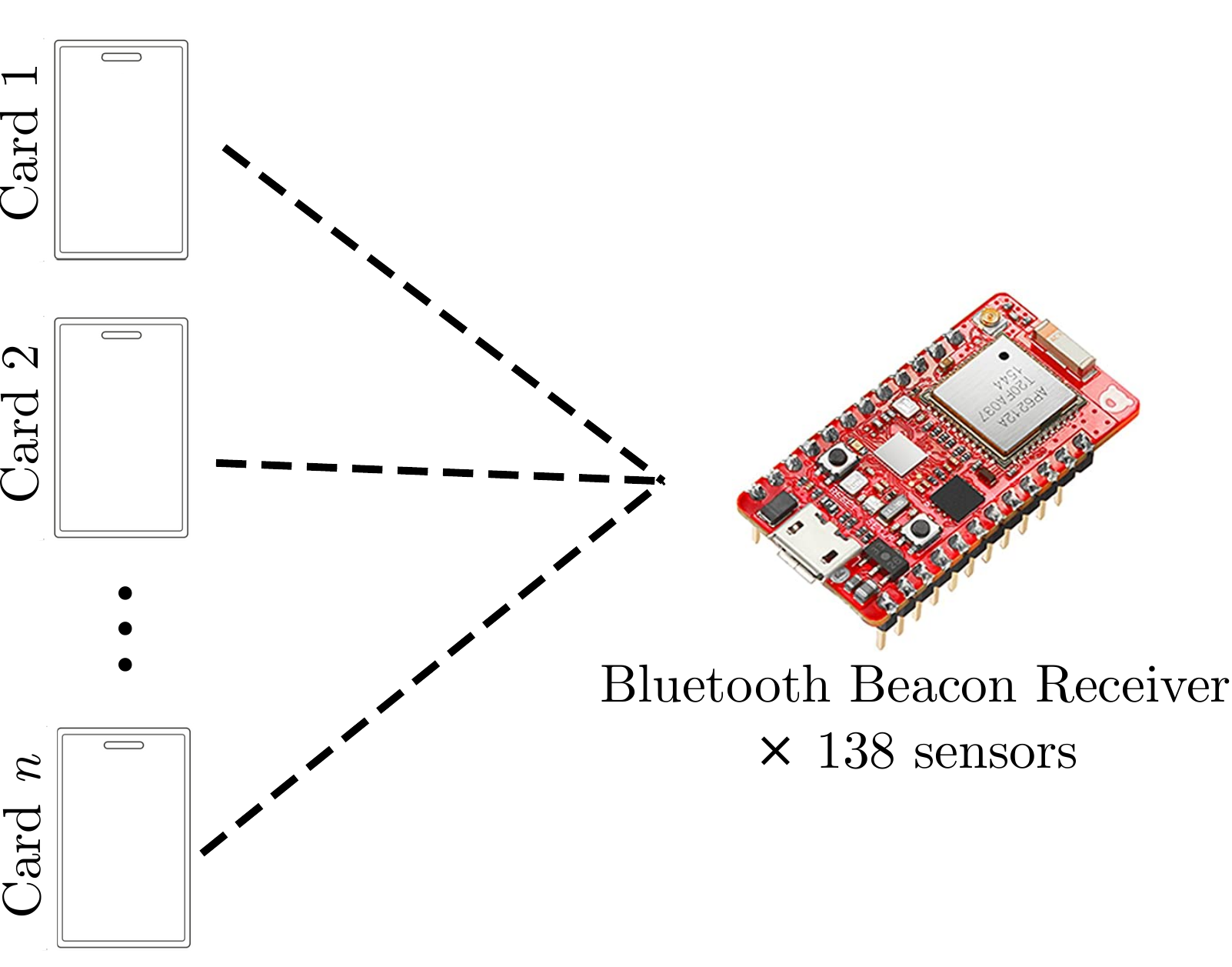}
	\caption{Card tags with Bluetooth beacon (Transmitter) and Redbear Duo (Receiver).}
	\vspace{0.2cm}
	\label{fig:hardware_card}
\end{subfigure}
\begin{subfigure}[sample2]{0.4\textwidth}	
	\centering \includegraphics[width=0.9\textwidth]{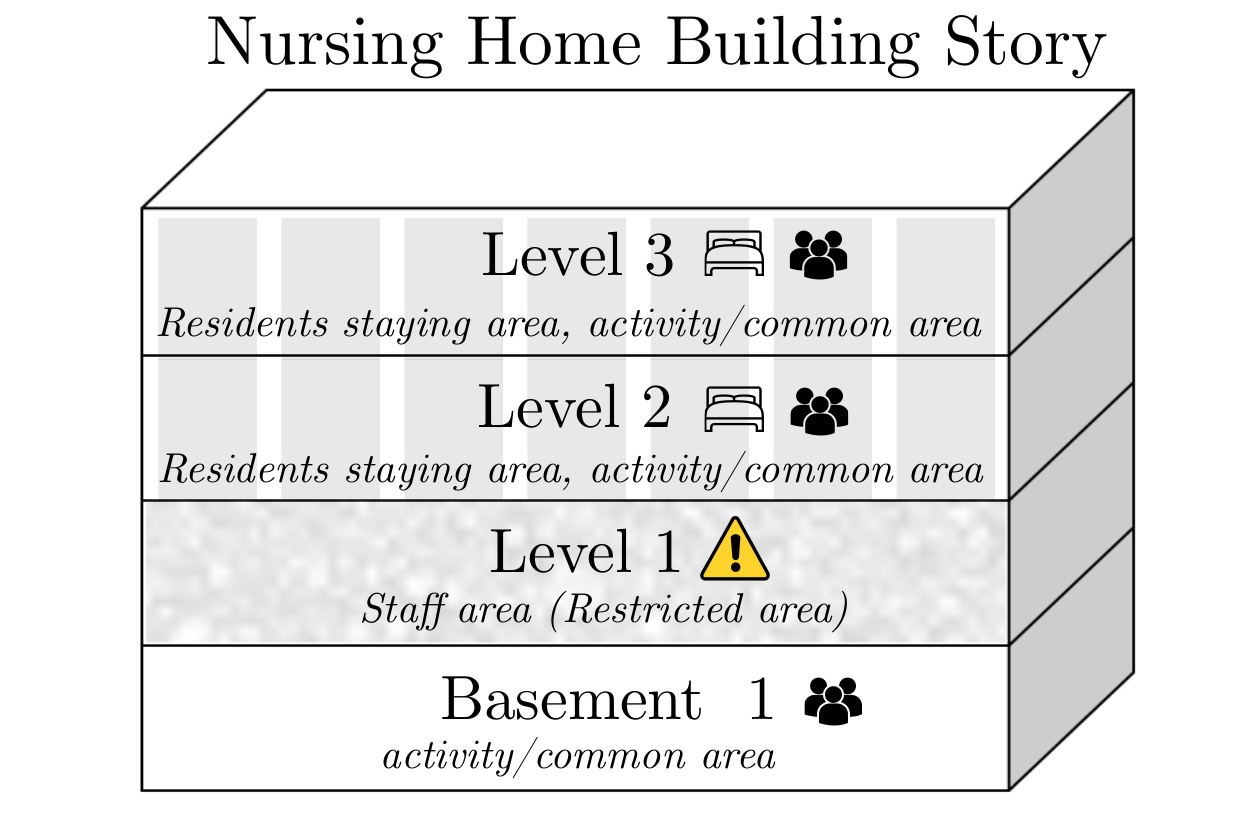}
	\vspace{-0.1cm}
	\caption{The nursing home building's floor level, which can be divided into 4 story. Note that Level 2 and 3 consists of residential staying area and common area, where Basement 1 only has common activity area. Level 1 is the nursing staff area, where elderly normally do not have access to that area.}
	\label{fig:hardware_building}
\end{subfigure}
\vspace{-0.2cm}
\caption{Hardware used to setup the monitoring system.}
\label{fig:hardware_all}
\vspace{-0.32cm}
\end{figure}

%%%%%%%%%%%%%%%%%%%%%%%%%%%%%%%%%%%%%%%%%%%%%%%%%%%%%%%%%%%%%%%%%%%%%%%%%%%%%%%%%%%%%%%%%%%%%%%%%%%%%%%%%%
\section{System Design}
\label{sec:systemDesign}

In this paper, we focus on the location-driven monitoring techniques since the resident of the nursing home stay within the premises.
Therefore, building-wide monitoring is more desirable in our studies. 
Given that collecting ground-truth appears to be impossible in our problem, the insights exploration approach and unsupervised machine learning method is more appropriate.
In this section, the overall system design of the deviated \revtwo{activity behavior} detection system is presented followed by the data specification and data preprocessing steps.
\vspace{-0.3cm}

%%%%%%%%%%%%%%%%%%%%%%%%%%%%%%%%%%%%%%%%%%%%%%%%%%%%%%%%%%%%%%%%%%%%%%%%%%%%%%%%%%%%%%%%%%%%%%%%%%%%%%%%%%
\subsection{Hardware Setup}
The proposed hardware setup comprises of two crucial components installed in the nursing home, which are card tags with BLE beacons and beacon receiver.
The Bluetooth beacon card model is shown in Fig.~\ref{fig:hardware_card}, which is capable of transmitting beacon every 1000ms using Nordic nRF52 chip with a range of up to $40$ meters.
\begin{algorithm}[H]	
\caption{Location Detection Algorithm}
\label{alg:decideLocation}
\fontsize{8pt}{8pt}\selectfont
\KwData{\textit{mac\_Address\_List, RSSI\_List, loc\_ID}}
\KwResult{\textit{user\_list, location\_list, timeStamp}}
\BlankLine 
\textbf{function} listenData()  \BlankLine \vspace{-0.11cm}
\Indp 1. Perform detection cycle as follows:\BlankLine \vspace{-0.11cm} 	\For{detection cycle from $1 $ to $5$}{
	\While{timer $< 3$ sec}{
		\textit{loc\_ID, RSSI\_List}  \BlankLine \vspace{-0.11cm} %$\leftarrow$  listen MQTT broadcast data
	}
	Filter \textit{mac\_Address\_List} $\le-70$dBm \BlankLine \vspace{-0.11cm}
	\For{unique resident in \textit{loc\_ID}}{ 
		\textit{location\_list} = getHighestRSSI(\textit{loc\_ID}, \textit{RSSI\_List})\BlankLine \vspace{-0.11cm} 
	}
	Store the \textit{userlist} and \textit{location} \BlankLine \vspace{-0.11cm}
}
2. Compute the final location after detection cycle ended\BlankLine \vspace{-0.11cm}
\ForEach{unique resident in \textit{userlist} }{
	determine the location based on the strongest RSSI\BlankLine \vspace{-0.11cm} 
	\If{computeLocation(location) $\ne$ NULL}{
		\textit{final\_location} $\leftarrow$ computeLocation(location)}
	\Else{
		Retrieve \textit{last\_location} from database \BlankLine \vspace{-0.11cm}
		\textit{final\_location} $\leftarrow$ \textit{last\_location}}
}\BlankLine \vspace{-0.11cm} 
3. update the resident \textit{final\_location} and \textit{current\_timestamp}
\BlankLine
\BlankLine
\Indm
\textbf{function} getHighestRSSI(\textit{loc\_ID}, \textit{RSSIList}) \BlankLine \vspace{-0.11cm} 
\Indp \textit{last\_location} $\leftarrow$ \textit{initial\_location}\BlankLine \vspace{-0.11cm} 
\textit{last\_RSSI} $\leftarrow$ \textit{initial\_RSSI}\BlankLine \vspace{-0.11cm} 
\textit{index} $\leftarrow$ 0\BlankLine \vspace{-0.11cm} 
\ForEach{RSSI in RSSI\_List}{
	\If{RSSI $>$ Last\_RSSI}{
		\textit{last\_RSSI} $\leftarrow$ RSSI \BlankLine \vspace{-0.11cm} 
		\textit{last\_location} $\leftarrow$ \textit{loc\_ID}[index]}
	\textit{index} $\leftarrow$ \textit{index} $+1$ \BlankLine \vspace{-0.11cm} }
\textbf{return} \textit{last\_location}
\BlankLine
\BlankLine
\Indm
\textbf{function} computeLocation(\textit{locationList}) \BlankLine \vspace{-0.11cm} 
\Indp initialize \textit{location\_dict} \BlankLine \vspace{-0.11cm} 
\ForEach{location in locationList}{
	update \textit{location\_dict} count with location} 
\If{ max(count(\textit{location\_dict})) exists}{
	\textbf{return} \textit{location} in max(count(\textit{location\_dict}))}
\Else{\textbf{return} \textit{NULL}}
\end{algorithm}
It is equipped with a battery and capable of transmitting the data for up to 6 months without recharging. 
To provide sufficient coverage for the nursing home, each room has a beacon receiver installed and pick up the Bluetooth beacon transmitted.
To filter out the irrelevant Bluetooth devices, a unique identifier is assigned to each of the card tags. 
The building studied as illustrated in Fig.~\ref{fig:hardware_building} consists of 4 levels with the top 2 levels served as the residential area, while the lower 2 levels are the staff area and basement level 1.
Note that the residential area and basement level consists of a common and dining area, where the residents can interact and have activity together.
In total, there is over $138$ Bluetooth beacon receiver installed to provide sufficient coverage to monitor residents' \revtwo{activity behavior}.
There is a total of $50$ residents within the nursing home who are agreed to participate in this study.
The beacon receiver uses RedBear Duo as shown in Fig.~\ref{fig:hardware_card}, which later transmits the collected Bluetooth beacon list to the local server for further processing.

After the local server received the broadcast messages, it performs threshold filtering to remove any weak signal less than $-70$dBm. 
Subsequently, we compute the residents' stay location using Algorithm~\ref{alg:decideLocation}. 
The Algorithm~\ref{alg:decideLocation} undergoes a cycle of $15$ seconds to determine the location of the residents based on the strongest RSSI signal.
\revtwo{The complexity of the Algorithm~\ref{alg:decideLocation} is $O(R)$, which it depends on the number of RSSI signals received, $R$.}

\begin{figure}[h]
\centering
\includegraphics[width=0.45\textwidth]{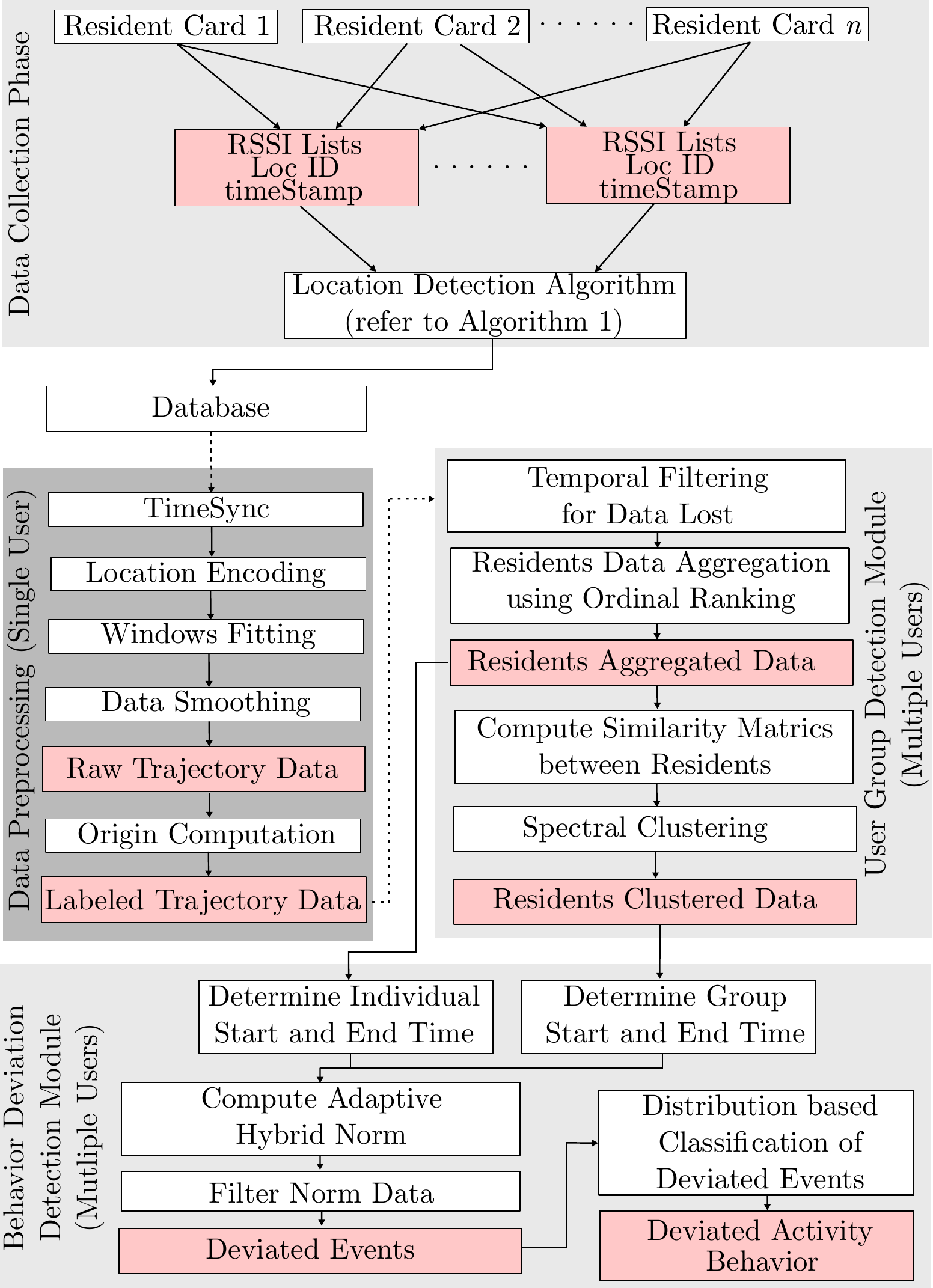} 
\caption{Data Processing Pipeline}
\label{fig:sys_archi}
\vspace{-0.56cm}
\end{figure}

\subsection{Data Specification and Processing pipeline}
After the residents' location data is stored in the database, we perform a series of processing onto the residents' trajectory data to extract the deviated \revtwo{activity behaviors}.
In this paper, 6 months of the residents' trajectory data is studied ranged from 01 Sep 2019 until 01 March 2020.
The overall data processing architecture is shown in Fig.~\ref{fig:sys_archi}.
Each resident undergoes the following preprocessing steps to extract their daily trajectory as well as residing room for further analysis. 
The preprocess module consists of the following steps, which are (1) time-sync, (2) location encoding, (3) windows fitting, and (4) data smoothing.  
The time-sync process is used to synchronize time for the data entry as previously proposed in~\cite{Lau2016Spatial}. 
Subsequently, the residents' locations are encoded into discrete numerical values for easier location representation.
Subsequently, windows fitting is performed to fit the location data into five-minute windows with the location with the longest duration denoted as stay location. 
Lastly, the data smoothing is performed to remove location with short duration stay, which could be a potential noise. 
After the preprocessing step, we would obtain a more structured residents' trajectory data.
%After preprocessing the data, we will obtain the trajectory for each respective resident. 

\revtwo{We have performed the complexity analysis on the system architecture to ensure the proposed system does not take ages to detect the activity behavior of a given nursing home resident. 
The computational complexity is $O(n^3)$, which the most time-consuming part is during clustering phase of the nursing home resident, $n$.
On the other hand, the space complexity of the proposed system is $O(n^2)$. }

Based on the trajectory, we compute the origin for each resident based on their longest stay duration and location from 11:00pm to 6:00am.
Note that origin denotes the room that a resident stayed in, while other rooms are denoted as private.
The encoded locations are divided into the following:
(a) origin level 2, (b) origin level 3, (c) private level 2, (d) private level 3, (e) public area basement level, (f) public area level 2, (g) public area level 3, and (h) restricted area.

In order to detect the different types of \revtwo{activity behavior} among the nursing home's residents, the clustering method utilizing a custom kernel is applied to compute similarity metrics across different elderly. 	
Based on the daily trajectory data, there are group activities among residents, where the residents are divided into multiple groups.
To find the common patterns among residents, the clustering approach is utilized to group residents with similar \revtwo{activity behavior}.
Further details of the clustering will be elaborated in Section~\ref{sec:groupDetection}.

After that, we want to study the deviated \revtwo{activity behavior} of the residents in the nursing home. 
Using the aggregated residents' data and cluster data, there are two types of norm data that can be computed, which are (1) individual, and (2) group norm data. 
Based on these two norms, a data fusion technique is used to generate daily hybrid norm for each resident and from there further extract each resident's deviated \revtwo{activity behavior}.
Subsequently, the deviated \revtwo{activity behavior} of the nursing home's residents is analyzed and categorized.

%%%%%%%%%%%%%%%%%%%%%%%%%%%%%%%%%%%%%%%%%%%%%%%%%%%%%%%%%%%%%%%%%%%%%%%%%%%%%%%%%%%%%%%%%%%%%%%%%%%%%%%%%%
\section{Group \revtwo{Activity Behavior} Clustering}
\label{sec:groupDetection}
%%%%%%%%%%%%%%%%%%%%%%%%%%%%%%%%%%%%%%%%%%%%%%%%%%%%%%%%%%%%%%%%%%%%%%%%%%%%%%%%%%%%%%%%%%%%%%%%%%%%%%%%%%
\vspace{-0.16cm}

\subsection{Clustering Algorithm}
In this subsection, we aim to study residents' \revtwo{activity behavior} by applying the clustering algorithm to group residents with similar trajectory patterns. 
To study the similarity between residents based on the location data, a custom similarity kernel is proposed to measure the resemblance between residents' \revtwo{activity behavior}. 
Each location is treated as categorical data and perform the windows sliding method to determine the similarity score between residents.
Let's denote each resident as $u_{i}$ in a nursing home, where the number of residents consists range of $i \in \{1,...,n\}$.
The resident $u_i$ is assigned a location $x_t$, which it consists of the spatial information $x$ and temporal information $t$ over the days $d$. 
This formulates the basic trajectory of residents in a nursing home.
By collecting the data over different days, a spatial-temporal matrix, $\textbf{X}_{i}$ consists of the temporal information $t$ can be denoted such as:
\begin{equation}
\textbf{X}_{i} = \begin{bmatrix}
	x_{1,1} & x_{2,1}  & ... & x_{1,t} \\ 
	x_{2,1} & x_{2,2}  & ... & x_{2,t} \\ 
	\vdots &\vdots   &  \ddots& \vdots\\ 
	x_{d,1} & x_{d,2}  & ... & x_{d,t} 
\end{bmatrix},
\label{eqn:vector}
\end{equation}
where the $d$ denotes number of days for the data collected and the $t$ denotes timestamp for each location $x$.
Based on the matrix $\textbf{X}_{i}$, the residents' location is aggregated to generate an individual pattern, $y_{i}$, which later will be used for clustering.
We consolidate the spatial-temporal matrix $\textbf{X}_{i}$ into a vector of $288$ samples, which each time-slot represents $5$ minutes interval of an encoded location.
The main motivation of choosing $288$ sample is we want to achieve between computing speed and data processing size.
This is computed using ordinal ranking~\cite{Agresti2010Analysis} based on the frequency of the location a resident stays throughout the data collection period. 
The aggregation trajectory for resident $u_i$ into $y_{i}$ can be shown in the following Algorithm~\ref{alg:aggTrajectory}:
\vspace{-0.12cm}
\begin{algorithm}[h]	
\caption{Trajectory Aggregate Function}
\label{alg:aggTrajectory}
\fontsize{8pt}{8pt}\selectfont
\KwData{trajectory, $X_{i}$} 
\KwResult{aggregated trajectory $y_{i}$}
\BlankLine 
1. Initialize vector $\textbf{V}$.  \BlankLine \vspace{-0.11cm}
2. Store location of all same timeslot \BlankLine \vspace{-0.11cm}
\For{$q$ from $0$ to $288$}{
	\ForEach{j}{
		\textbf{V}$[q] \leftarrow X_{j}$ 
}}
3. Initialize the aggregated trajectory $y_{i}$ \BlankLine \vspace{-0.11cm}
4. Perform aggregate function for each timeslot \BlankLine \vspace{-0.11cm}
\For{$q$ from $0$ to $288$}{
	Perform ordinal ranking for location based on the frequency of \textbf{V}$[q]$}
\end{algorithm}

After obtaining the residents' aggregated data, we calculate the similarity metrics between different residents for clustering purposes in order to generate a similar group. 
Therefore, the Weighted Windowed Overlap (WWO) similarity kernel is introduced to calculate the similarity between pairwise resident $u_{a}$ and $u_{b}$.
The WWO function $dist(u_{a}, u_{b})$ can be defined as following:
\begin{equation}
dist(u_{a}, u_{b}) = \frac{1}{t}\sum_{i=0}^{t} \frac{1}{2h+1}\left (\sum_{j=i-h}^{i+h} \left\{ {\begin{array}{*{20}{c}}
		1 & \text{if }y_{a,j} \ne  y_{b,j}\\
		0 & \text{if }y_{a,j} = y_{b,j}
\end{array}} \right. \right ),
\label{eqn:distKernel}
\end{equation}
where $\rho$ denotes window sliding parameter over the time $t$ by comparing the location of pairwise residents $(u_{a},u_{b})$.
Meanwhile, $h$ represents the threshold of the windows sliding mechanism.
In this paper, the threshold of $h$ is defined as $30$ minutes. 
Subsequently, the similarity calculation, $s$ can be calculated using the following equation:
\begin{equation}
s(u_a,u_b) = \textbf{W} \times \frac{1}{1 + dist(u_a, u_b)}
\label{eqn:similarityCalculation}
\end{equation}
where $W$ denotes the weight vector corresponding to the time of the day.

To ensure the effectiveness of computing similarity, we use a toy example of two different residents with the same origin but different \revtwo{activity behavior} patterns to study similarity measure as shown in Fig~\ref{fig:toyExample}(a) as follows:

\begin{figure}[ht]
\centering
\begin{tabular}{c}
	\includegraphics[width=0.47\textwidth]{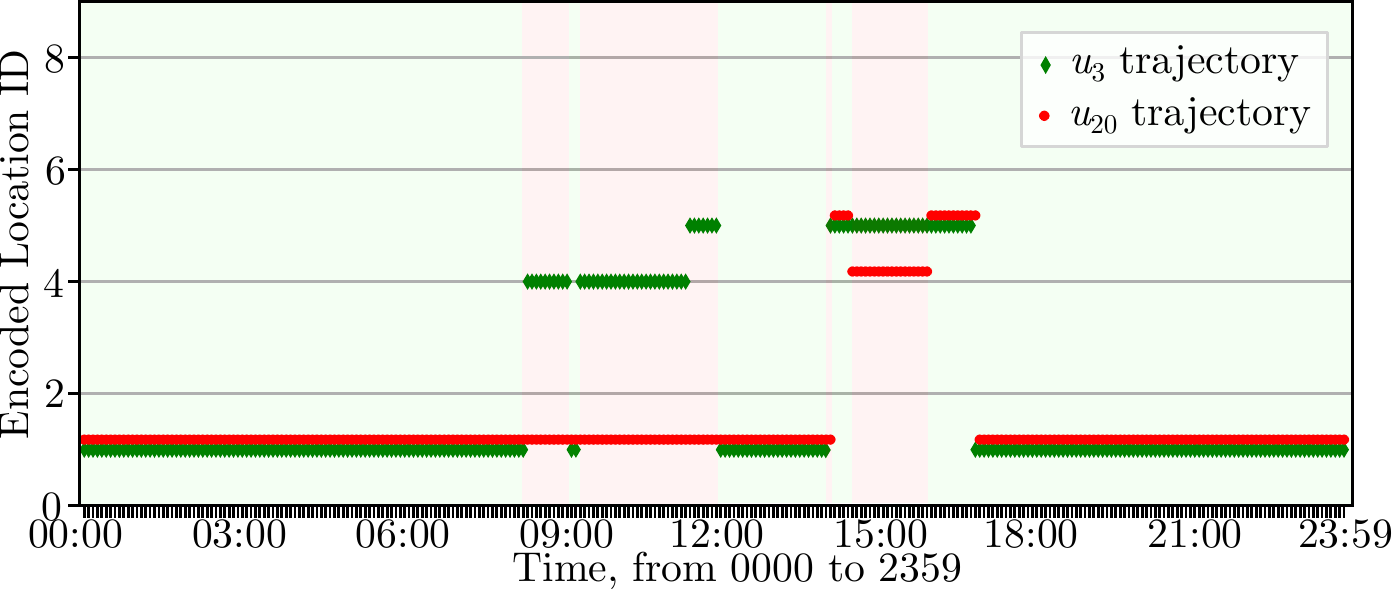} 
	\vspace{-0.1cm}\\
	\begin{tabular}{c}
		(a) Two trajectory examples (users pair($u_{3}$, $u_{20}$)) in \\
		encoded location for 11 Nov and 25 Oct 2019.
	\end{tabular}\\
	\includegraphics[width=0.47\textwidth]{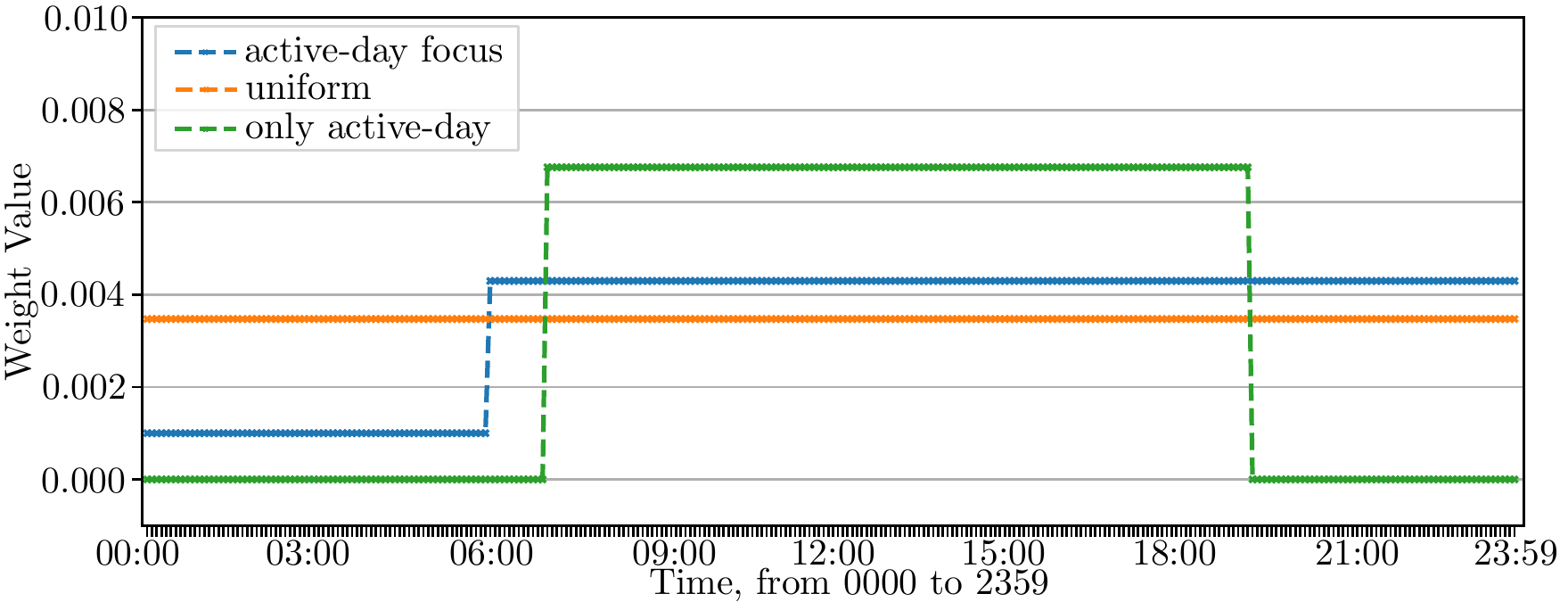} 
	\vspace{-0.1cm} \\
	(b) Weights used for calculating the similarity function
\end{tabular}
\vspace{-0.2cm} \\
\caption{Toy examples for calculating similarity metric and custom weight for similarity calculation.}
\label{fig:toyExample}
\end{figure}

Three different weight vectors are being considered when computing the similarity kernels for resident $u_{a}$ and $u_{b}$ using WWO, which are (1) uniform, (2) active-day focus, and (3) only active-day. 
The illustration of varying weight is presented in Fig~\ref{fig:toyExample}(b).
These three weight vectors try to highlight different temporal parts of the day and weight is adjusted accordingly.
For instance, active-day focus emphasizes the active period of the day (06:00am - 11:59pm), where only active-day only consider the active period (07:00am - 08:00pm).
Note that the total value of weight is $1.0$ and is allocated across $288$ vectors depending on the weight characteristic.
We choose one of the weights depending on the emphasis of the outcome of the desired similarity measurement.
Using some common similarity measurement as introduced in~\cite{Boriah2008Similarity}, a comparison of the similarity measurement is shown in Table~\ref{tbl:similarityComparison}.

\begin{table}[h!]
\caption{Comparison of Similarity Measurement}
\centering
\begin{tabular}{@{}l|c@{}}
	\toprule
	\multicolumn{1}{c|}{Similarity Measurement Method} & Similarity Score \\ \midrule
	WWO (active-day focus) & 0.7250 \\
	WWO (uniform) & 0.7778 \\
	WWO (only active-day) & 0.5675 \\
	Overlap~\cite{Stanfill1986Memory} & 0.8182 \\
	Eskin~\cite{Eskin2002Geometric} & 0.9259 \\
	Goodall~\cite{Goodall1966New} & 0.9254 \\ \bottomrule
	%		WWO (active-day focus) & 0.4415 \\
	%		WWO (uniform) & 0.5486 \\
	%		WWO (only active-day) & 0.1791 \\
	%		Overlap~\cite{Stanfill1986Memory} & 0.6889 \\
	%		Eskin~\cite{Eskin2002Geometric} & 0.8494 \\
	%		Goodall~\cite{Goodall1966New} & 1.0000 \\ \bottomrule
\end{tabular}
\label{tbl:similarityComparison}
\end{table}

Based on the observation, the variation of overlap methods has similar results ranging from $0.7250$ to $0.8182$, where the similarity score of the WWO with only active-day is $0.5675$.
Meanwhile, the other methods (Eskin and Goodall) have a higher similarity score despite the visualization of two residents' trajectory in Fig.~\ref{fig:toyExample} shows different patterns. 
Ideally, the overlap method presents the most straightforward method of computing the similarity score, which is roughly around $0.8182$.
However, the overlap method only highlights the similarity between two users during night time (08:00pm - 06:00am), which represents a large amount of time when users are inactive.
Thus, after considering the different weightage similarity scores, active-day focus weight is more desirable for computing the similarity metric, where it emphasizes a more active period on the day.

By iterating the similarity score through different pairwise residents, we can obtain the similarity matrix, $\textbf{A}$ as such:

\begin{equation}
\textbf{A} = \left[ \text{\footnotesize ${\begin{array}{*{20}{c}}
			%	{0}& {{s_{1,2}}}& \ldots& {{s_{1,n-1}}} &{{s_{1,n}}}\\
			%	{{s_{2,1}}}&{0}& \ldots& {{s_{2,n-1}}} &{{s_{2,n}}}\\
			%	\vdots & \vdots & \ddots& \vdots  & \vdots \\
			%	{{s_{n-1,1}}}&{{s_{n-1,2}}}& \cdots &0 &{{s_{n-1,n}}}\\
			%	{{s_{n,1}}}&{{s_{n,2}}}& \cdots &{{s_{n,n-1}}} &{0}\\
			{0}& {{s(u_1, u_2)}}& \ldots& {{s(u_1, u_{n-1})}} &{{s(u_1, u_{n})}}\\
			{{s(u_2, u_{1})}}&{0}& \ldots& {{s(u_2, u_{n-1})}} &{{s(u_2, u_{n})}}\\
			\vdots & \vdots & \ddots& \vdots  & \vdots \\
			{{s(u_{n-1}, u_{1})}}&{{s(u_{n-1}, u_{2})}}& \cdots &0 &{{s(u_{n-1}, u_{n})}}\\
			{{s(u_{n}, u_{1})}}&{{s(u_{n}, u_{2})}}& \cdots &{{s(u_{n}, u_{n-1})}} &{0}\\
	\end{array}}$} \right]
\label{eqn:similaritymatrix}
\end{equation}
In order to compute the laplacian matrix, degree matrix, $D$ can be computed as follows:
\begin{equation}
D = \sum_{i}^{n} \left\{ {\begin{array}{*{20}{c}}
		1 & \textbf{A}_{i,i} \ge 0  \\
		0 & \text{ otherwise}
\end{array}} \right. ,
\label{eqn:degreeMatrix}
\end{equation}
where it represents a non-zero affinity matrix.

Next, using Eqn.~\ref{eqn:similaritymatrix} and Eqn~\ref{eqn:degreeMatrix}, the normalized Laplacian matrix, $\textbf{L}$ can be generated using following equation:
\begin{equation}
\textbf{L}=\textbf{I} - \textbf{D}^{-1/2} \textbf{A} \textbf{D}^{-1/2} ,
\label{eqn:LaplacianMatrix}
\end{equation}
where $I$ denotes the identity matrix.

\begin{algorithm}[h]	
\caption{Spectral Clustering Algorithm}
\label{alg:stayPointClustering}
\fontsize{8pt}{8pt}\selectfont
\KwData{Spatio-temporal Matrix, $\textbf{X}$}
\KwResult{Cluster List, $C$}
\BlankLine 
1. Aggregate data into daily windows data. \BlankLine \vspace{-0.11cm}
2. Calculate the affinity matrix, $\textbf{A}$ as follows: \BlankLine \vspace{-0.11cm}
\For{$i \leftarrow 1 $ to $n$}{
	\For{$j \leftarrow 1 $ to $n$}{
		calculate the similarity metric using Eqn.~\ref{eqn:distKernel} for each $i$ and $j$ resident. \BlankLine \vspace{-0.11cm} 
	}
}
3. Compute the degree matrix, $\textbf{D}$ using Eqn.~\ref{eqn:degreeMatrix}. \BlankLine \vspace{-0.11cm} 
4. Calculate the Laplacian Matrix, $\textbf{L}$ using Eqn.~\ref{eqn:LaplacianMatrix}. \BlankLine \vspace{-0.11cm} 
5. Calculate the Eigenvector, $\textbf{U}$. \BlankLine \vspace{-0.11cm} 
6. Determine $k$ value based on sum of squared distance (SSD). \BlankLine \vspace{-0.11cm} 
7. Perform  $k$-means and obtain cluster list, $C$. \BlankLine \vspace{-0.11cm} 
return $C$.
\end{algorithm}

Subsequently, we compute the $k$ generalized eigenvectors using the normalized Laplacian matrix, $\textbf{L}$ as follows:
\begin{equation}
\textbf{L}u = \lambda \textbf{D}u
\end{equation}
where vector, $u$ is the calculated using the smallest $k$ eigenvalue.
By combining the afore-mentioned equation, we formulate the spectral clustering in the Algorithm~\ref{alg:stayPointClustering}. 
\revtwo{The computational complexity of the spectral clustering is $O(n^3)$, which is the most time consuming part of the proposed system}.
Based on the proposed algorithm, we perform group detection and show the result in next sub-section.

\subsection{Results and Validation}
\begin{figure}[h]
\centering
\includegraphics[width=0.45\textwidth]{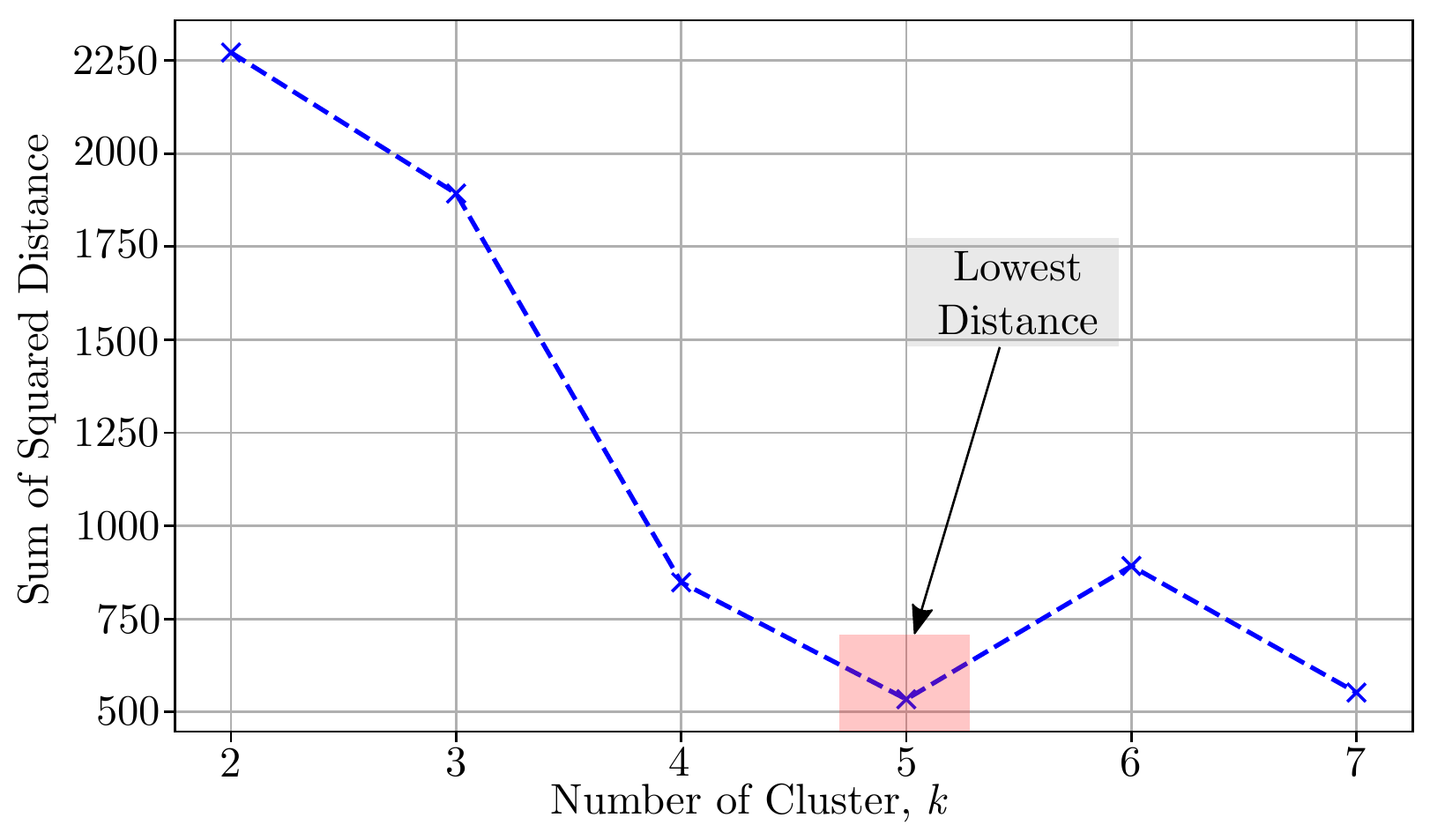} 
\vspace{-0.2cm}
\caption{SSD for different number of $k$.}
\label{fig:SSDclusterScore}
\end{figure}

In order to study the optimal $k$ for deciding the number of clusters in the nursing home, we apply the sum of squared distance (SSD) for different number of $k$ values. 
The SSD can be defined as follows:
\begin{equation}
\sum^{k} \sum_{a=1}^{n}\sum_{b=1}^{a-1}{(dist(u_a, u_b))^2},
\end{equation}
where it utilizes $dist()$ from Eqn.~\ref{eqn:distKernel}. 
It calculates the summation of squared distance for different $k$ values in clustering, which lower value indicates higher similarity between residents in the same cluster. 
Ideally, we want to find clusters within the range of $2$ to $7$ clusters out from $50$ residents.
The SSD result is presented in Fig~\ref{fig:SSDclusterScore}.

\begin{figure*}[h]
\centering
\begin{tabular}{cc}
	\includegraphics[width=0.5\textwidth]{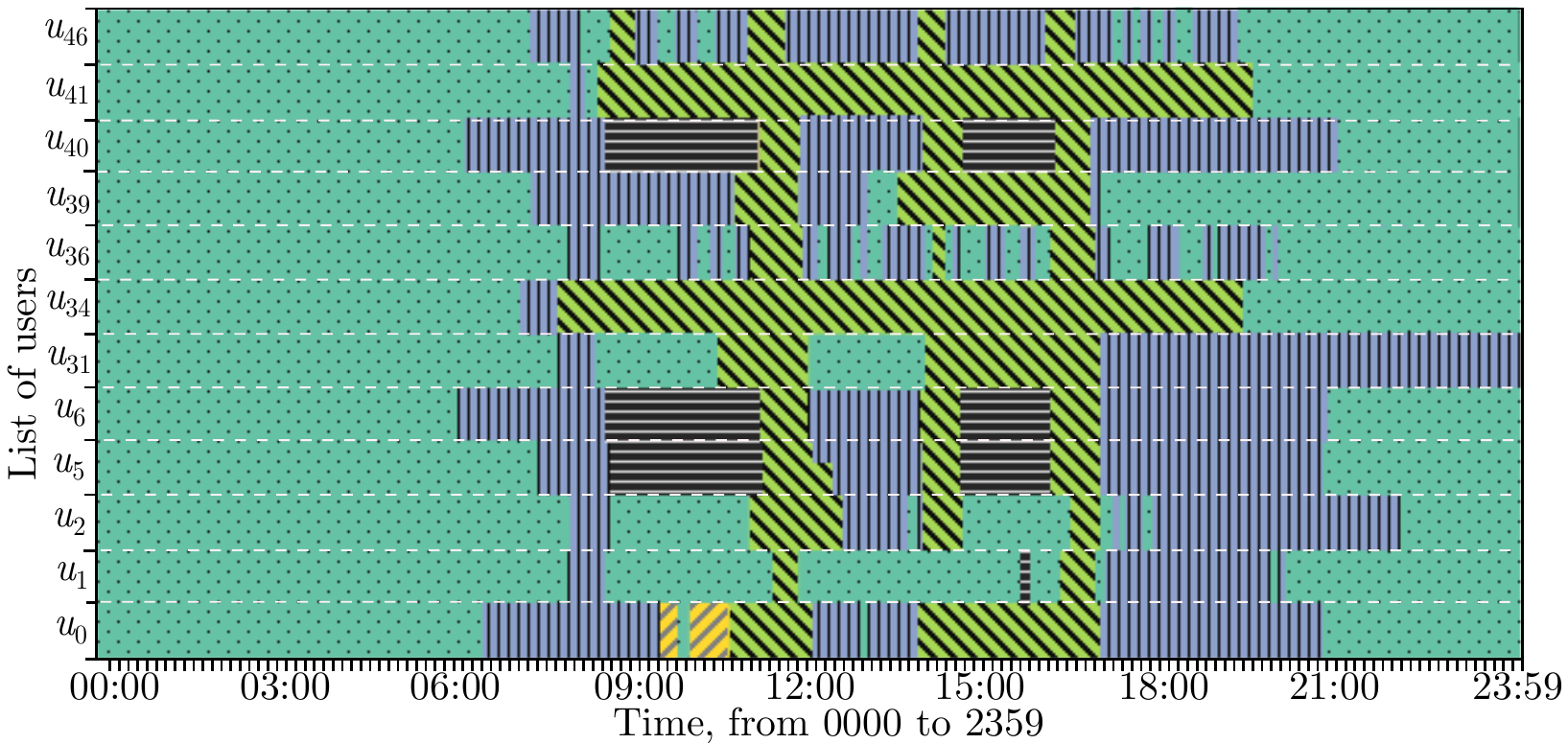} 	&  
	\includegraphics[width=0.5\textwidth]{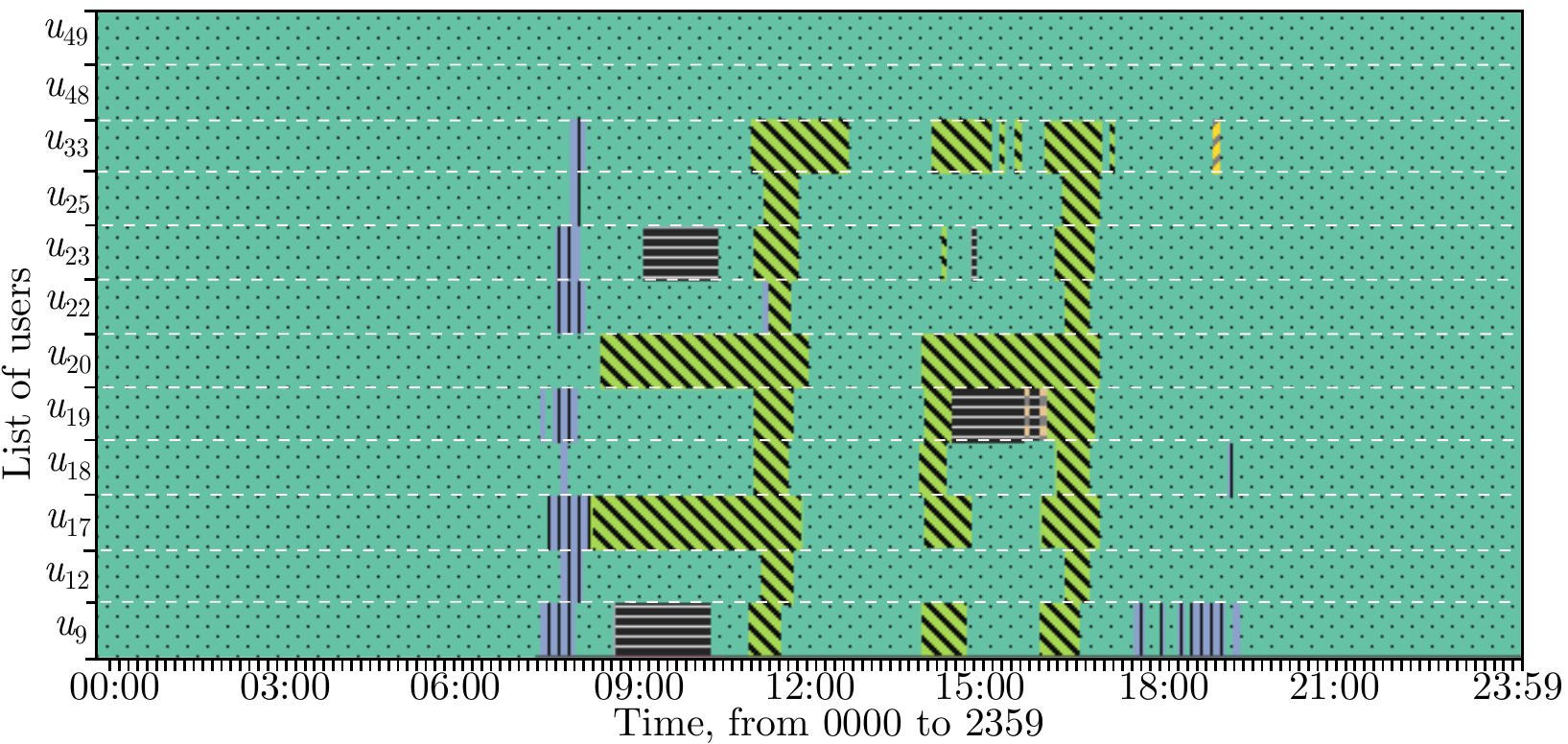} \\
	(a) Cluster 1 & (b) Cluster 2 \\
	\includegraphics[width=0.5\textwidth]{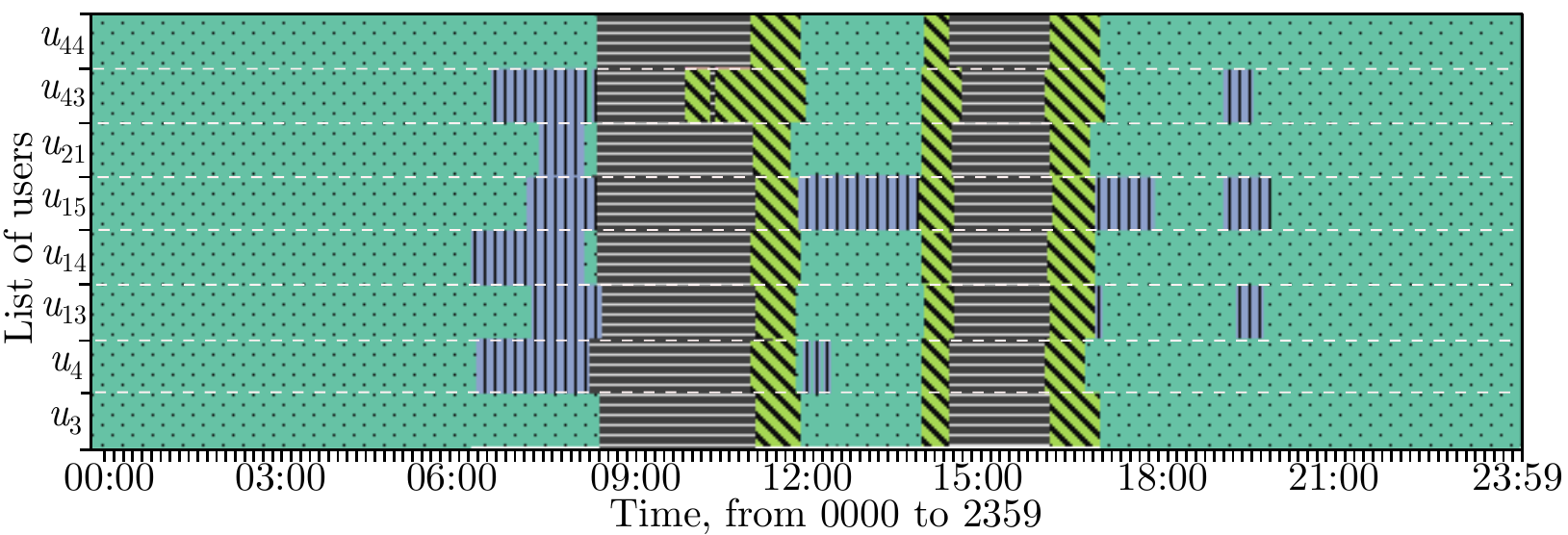} 	&  
	\includegraphics[width=0.5\textwidth]{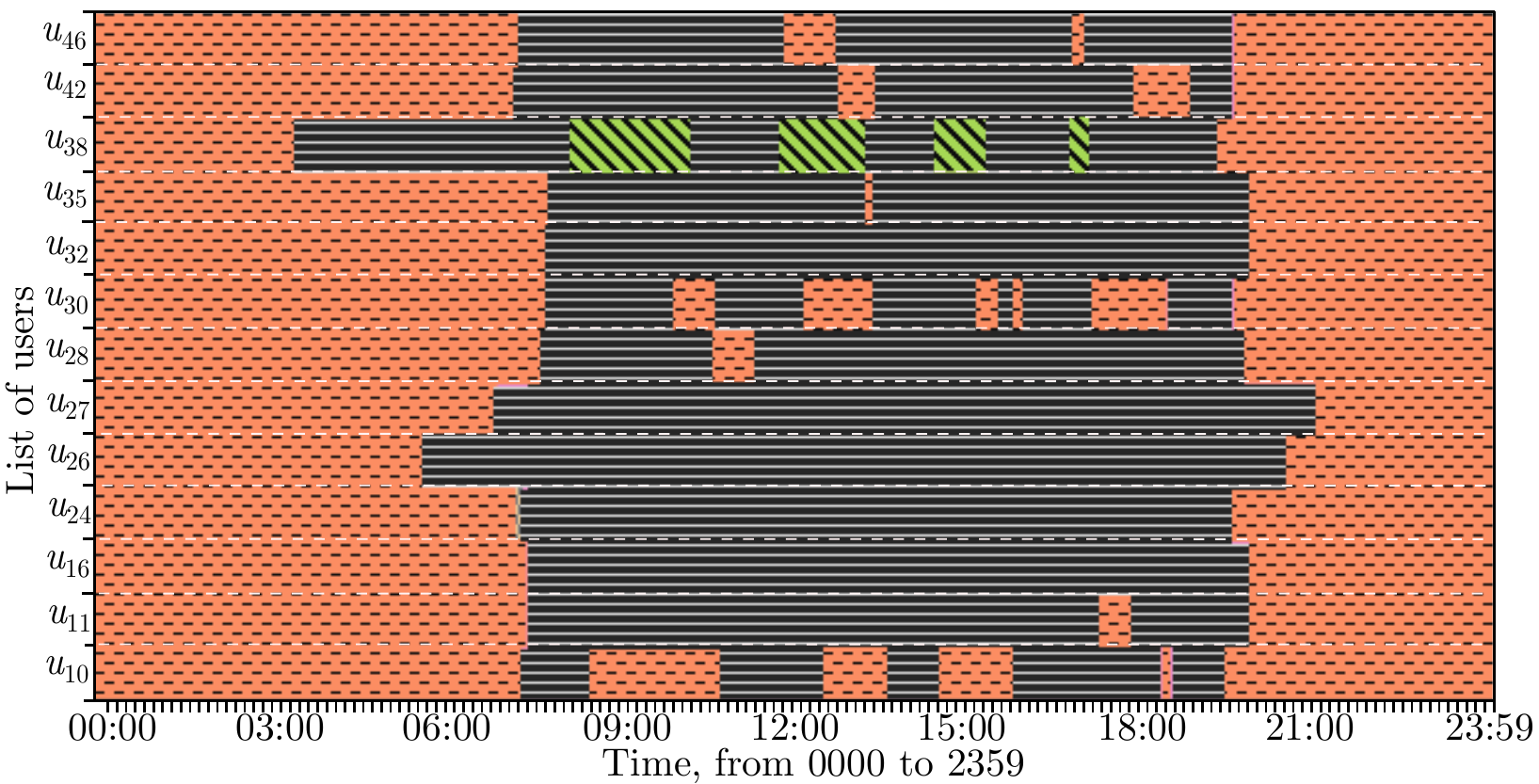} \\
	(c) Cluster 3 & (d) Cluster 4 \\
	\includegraphics[width=0.5\textwidth]{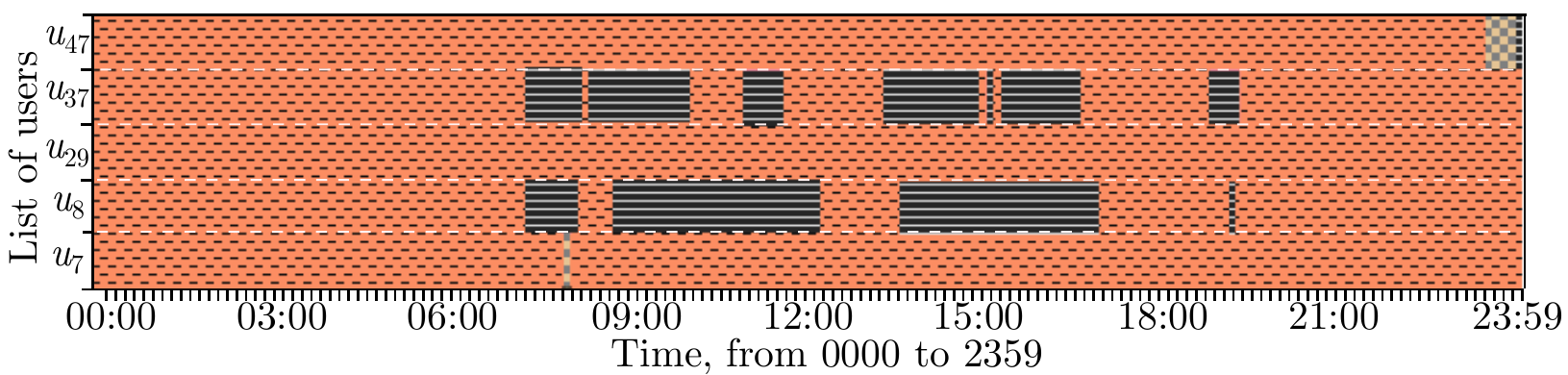} 	&
	\includegraphics[width=0.45\textwidth]{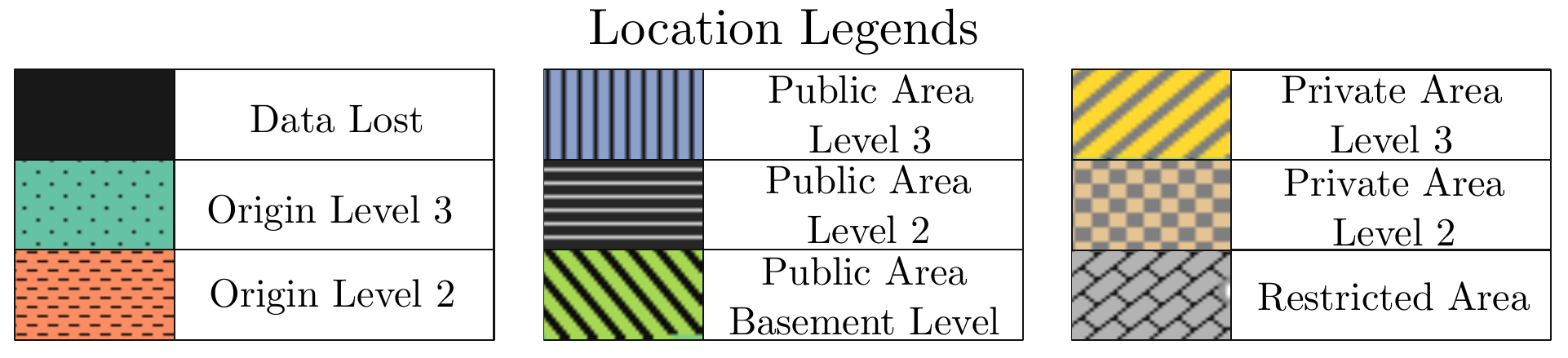} \\		
	(e) Cluster 5 & 
\end{tabular}
\vspace{-0.2cm}
\caption{Group Clustering Result based on $k$=5.}
\label{fig:clusterResult}
\end{figure*}

Based on observation, $k$=$5$ is an ideal number for the clusters as the SSD value is the lowest compared to other choices of $k$.
Using $k$=$5$, the clustering algorithm is formulated as shown in Algorithm~\ref{alg:stayPointClustering}, and the clustering result is presented in Fig.~\ref{fig:clusterResult}. 

We observed clusters are separated by building level and have their own characteristic.
The level 3 residents generally can be divided into three different groups, where level 2 can be divided into 2 different groups. 
The residents from Cluster 1 tend to have a longer visit duration at level 3 public area compared to Cluster 2 and Cluster 3. 
Meanwhile, resident $u_{15}$ in Cluster 3 spends his/her lunchtime in the level 3 public area instead of the basement public area.

Meanwhile, residents from Cluster 2 tend to spend less time in public places, where Cluster 3's residents spend more time in level 2 public areas. 
In Cluster 2, resident $u_{48}$ and resident $u_{49}$ spend time in their room without going to public spaces, which is quite peculiar. 
Level 2's residents in Cluster 4 mostly spend their time in the public area, while residents in Cluster 5 spend less time in the public area and spend most of their time in their respective origin.

Based on these 5 clusters, we will generate daily activity based on their group to compute their daily \revtwo{activity behavior}.
Further details of fusing individual \revtwo{activity behavior} and group \revtwo{activity behavior} are elaborated in the next section.

%\pagebreak 
%%%%%%%%%%%%%%%%%%%%%%%%%%%%%%%%%%%%%%%%%%%%%%%%%%%%%%%%%%%%%%%%%%%%%%%%%%%%%%%%%%%%%%%%%%%%%%%%%%%%%%%%%%
\section{Residents' Deviated \revtwo{Activity Behavior} Study}
\label{sec:irregularStudy}
%%%%%%%%%%%%%%%%%%%%%%%%%%%%%%%%%%%%%%%%%%%%%%%%%%%%%%%%%%%%%%%%%%%%%%%%%%%%%%%%%%%%%%%%%%%%%%%%%%%%%%%%%%
\subsection{Residents' Hybrid Norm Computation}
Despite there is group \revtwo{activity behavior} among the nursing home's residents in their daily routine, there exist cases of some residents who behave differently than the others.
The trajectory of the residents may vary day to day depending on the group schedule for the day.
Therefore, it is crucial to consider the residents' groups daily trajectory to provide additional information to generate daily norms for studying potential deviated \revtwo{activity behavior}.
To show an example of deviated \revtwo{activity behavior} detection, we use resident $u_{21}$'s norm and the group norm (Cluster 3) as an example to demonstrate the working of extracting the deviated locations from the data.
The example of input data and hybrid norm generation are shown in following Fig.~\ref{fig:exampleHybridExtraction}:

\begin{figure*}[ht]
\centering
\includegraphics[width=0.90\textwidth]{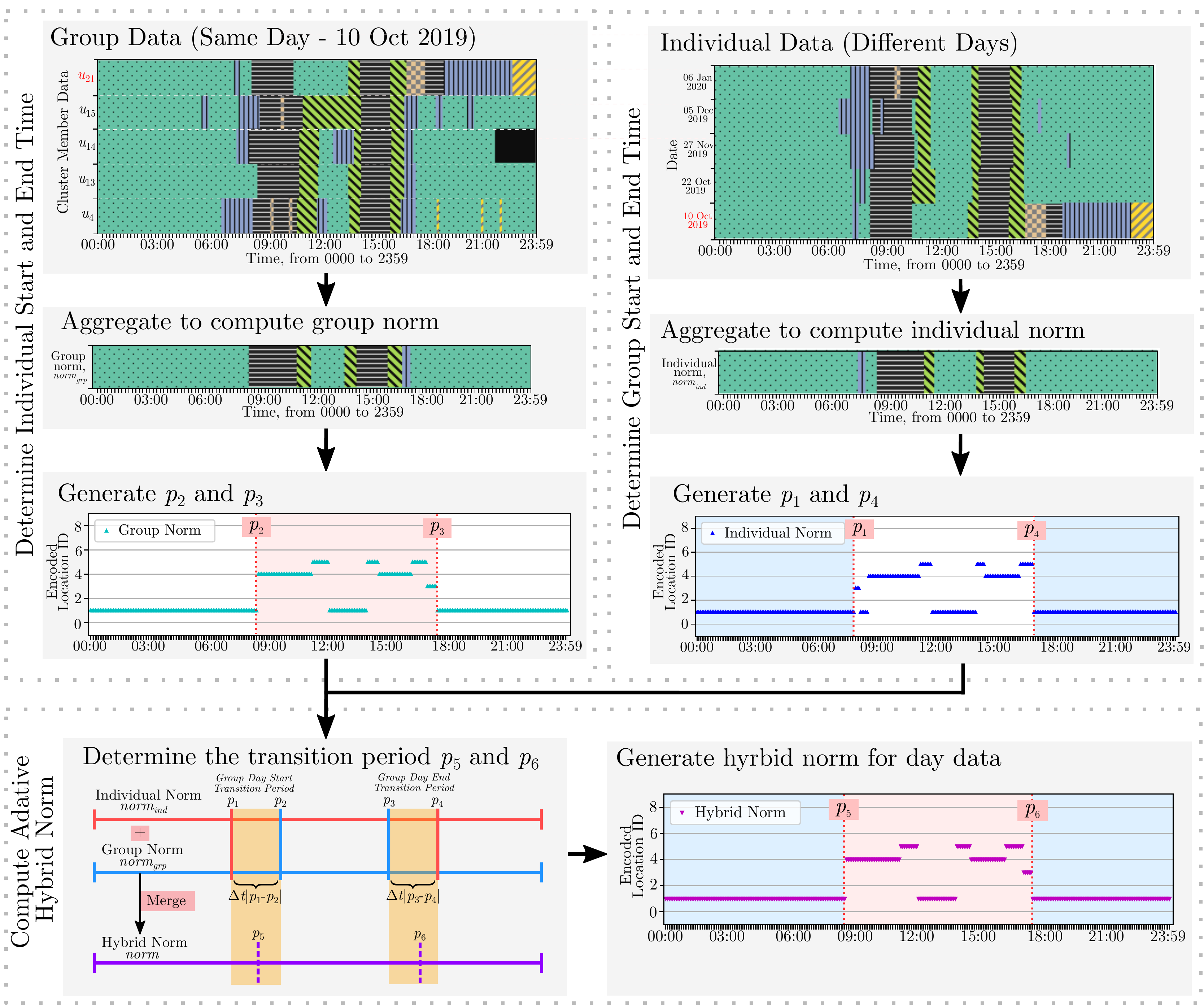} 
\vspace{-0.2cm}
\caption{Toy example for extracting deviated \revtwo{activity behavior} using resident $u_{21}$'s trajectory data.}
\label{fig:exampleHybridExtraction}
\vspace{-0.16cm}
\end{figure*}

Using the example, we extract two different norms from the resident $u_{21}$ to generate a hybrid norm for 10 October 2019.
Both norms used as data input to generate a hybrid norm can be defined as follows:
\begin{itemize}
\item \textit{Individual Norm}, $norm_{ind}$ -
The individual norm represents the regular pattern of the resident over the data collection period. The regular pattern is generated through the aggregated of the resident's valid data, which can be obtained using the user group detection module's Algorithm~\ref{alg:aggTrajectory}. Note that the individual norm is obtained using aggregated trajectory $y_i$ for each resident, $u_i$.
\item \textit{Group Norm}, $norm_{grp}$ - 
Group norm denotes the regular pattern of all clustered residents obtained using the clustering algorithm in the previous section. Note that this pattern changes every day as a nursing home have different activities/events arranged during the active hour. Therefore, $norm_{grp}$ is calculated every day for each cluster.
\end{itemize}

Utilizing these norms, we propose a data fusion method to generate a hybrid norm $norm$. 
The main challenge here is to address the transition method between $norm_{ind}$ and $norm_{grp}$, where overlapping between two norms is possible when combining both norms.

\begin{algorithm}[h]	
\caption{Generating Adaptive Hybrid Norm}
\label{alg:adaptiveHybridNorm}
\fontsize{8pt}{8pt}\selectfont
\KwData{\textit{group\_label}, Day Data $\textbf{X}_d$, origin $x_{ori}$, threshold $h$}
\KwResult{Hybrid\_Norm $norm_{d,i}$}
\BlankLine 
\textbf{function} generateHybridNorm()  \BlankLine \vspace{-0.11cm}
\Indp 1. Obtain $norm_{ind}$ using Algorithm~\ref{alg:aggTrajectory}. \BlankLine \vspace{-0.11cm} 
2. Obtain $norm_{grp}$ using \textit{getGroupNorm}(\textit{group\_label}, $\textbf{X}_d$). \BlankLine \vspace{-0.11cm}
3. Compute $p1$ and $p4$ using \textit{determineDayStartEnd}($norm_{ind}$, $x_{ori}$, h).\BlankLine \vspace{-0.11cm}
4. Compute $p2$ and $p3$ using \textit{determineDayStartEnd}($norm_{grp}$, $x_{ori}$, h).\BlankLine \vspace{-0.11cm}
5. Compute $p5$ and $p6$ using Eqn.~\ref{eqn:startTransTime} and Eqn.~\ref{eqn:endTransTime}.\BlankLine \vspace{-0.11cm}
6. Merge the data to become adaptive hybrid norm using Eqn.~\ref{eqn:mergeHybridNorm} \BlankLine \vspace{-0.11cm} 
\BlankLine
\BlankLine
\Indm
\textbf{function} getGroupNorm(\textit{group\_label}, $\textbf{X}_{d}$) \BlankLine \vspace{-0.11cm} 
\Indp 1. Initialize the same\_group\_dictionary \BlankLine \vspace{-0.11cm} 
2. Add the residents' data, $X$ if they are from the same group\BlankLine \vspace{-0.11cm} 
\ForEach{$\textbf{X}_{d,j}$}{\BlankLine \vspace{-0.11cm} 
	\If{resident $\in$ group\_label}{same\_group\_dictionary append $\textbf{X}_{d,j}$}} \BlankLine \vspace{-0.11cm} 
3. Check whether the same\_group\_dictionary has more than $2$ entries. \BlankLine \vspace{-0.11cm} 
\If{length( same\_group\_dictionary) $> 2$}{
	\textbf{return} \textit{aggregate same\_group\_dictionary using Algorithm~\ref{alg:aggTrajectory}}
}
\Else{\textbf{return} \textit{NULL}}
\BlankLine
\BlankLine
\Indm
\textbf{function} determineDayStartEnd($X$, $x_{ori}$, $h$) \BlankLine \vspace{-0.11cm} 
\Indp 
1. Compute day start pointer \BlankLine \vspace{-0.11cm}
\For{$t$, $0$ to $144$}{
	\If{$x_{t} \ne  x_{ori}$ for $h$ interval}{break}
}
2. Store the value, \textit{DayStart} $\leftarrow t$ \BlankLine \vspace{-0.11cm} 
3. Compute day end pointer  \BlankLine \vspace{-0.11cm}
\For{$t$, $288$ to $144$}{
	\If{$x_{t} \ne  x_{ori}$ for $h$ interval}{break}
}
4. Store the value, \textit{DayEnd} $\leftarrow t$ \BlankLine \vspace{-0.11cm} 
\textbf{return} \textit{DayStart, DayEnd}						
\end{algorithm}

There are two transition periods in the day of study, which are the group activity start transition period and group activity end transition period.
One needs to decide the exact timing for transition based on the starting period of the group using $p_2$ and ending period $p_3$.
Moreover, similar consideration applied for the ending time of the private period, $p_1$ and the starting period of private period $p_4$.
The $p_5$ denotes the starting of the group activity, while $p_6$ represents the group activity ending period. 
To compute the transition period $[p_5, p_6]$ using $[p_1, p_2, p_3, p_4]$, the following Eqn.~\ref{eqn:startTransTime} is used and \ref{eqn:endTransTime} to decide the transition period for group starting and group ending period:
\begin{equation}
p_5 = 
\left\{\begin{matrix}
	p_1 & \text{if } \Delta t |p_1-p_2| \le h \text{ or } \Delta t (p_1-p_2) \ge 0\\ 
	p_2 & \text{ otherwise }
\end{matrix}\right. ,
\label{eqn:startTransTime}
\end{equation}
\begin{equation}
p_6 = 
\left\{\begin{matrix}
	p_4 & \text{if } \Delta t |p_3-p_4| \le h \text{ or } \Delta t (p_3-p_4) < 0\\ 
	p_3 & \text{ otherwise }
\end{matrix}\right. ,
\label{eqn:endTransTime}
\end{equation}
where $h$ denotes the time gap limit between 2 time slots such as $(p_1, p_2)$ and $(p_3, p_4)$. 
Note that if the time overlap such as $\Delta t (p_1-p_2) \ge 0$ or $\Delta t (p_3-p_4) < 0$, the earliest time is considered as transition period as default overlapping time.

After deciding the transition period, we compute the hybrid norm, $norm$ by combining the individual and group norm of the nursing home residents.
The merging process between $norm_{ind}$ and $norm_{grp}$ can be formulated as follows:
\begin{equation}
\begin{matrix}
	norm_{d,i}[t=0:p_5] & \leftarrow norm_{ind}[t=0:p_5]\\ 
	norm_{d,i}[p_5:p_6] & \leftarrow norm_{grp}[p_5:p_6]\\ 
	norm_{d,i}[p_6:t=288] & \leftarrow norm_{ind}[t=p_6:288]
\end{matrix}
\label{eqn:mergeHybridNorm}
\end{equation}	
where the period between $t$=$0 \text{ to } p_5$ and $t$=$p_6 \text{ to } 288$  uses norm from $norm_{ind}$. 
The $p_5 \text{ to } p_6$ denotes the group time.
In addition, $d$ represents the particular day we are studying, while $i$ denotes each of the residents in the nursing home. 
\revtwo{By combining the aforementioned equations, the hybrid norm $norm_{d,i}$ for each resident $u_i$ can be computed using Algorithm~\ref{alg:adaptiveHybridNorm} and the computational complexity is $O(n^2)$.}

\subsection{Residents' Deviated \revtwo{Activity Behavior} Identification}
After computing the adaptive hybrid norm for each user, we extract the deviated locations using a filtering function based on the norm and daily input data, $\textbf{X}_d$.
Since the encoded location is categorical data, the binary comparison method is adopted to compare whether the input data is the same as the generated adaptive hybrid norm. 
The Eqn.~\ref{eqn:filterTechnique} describe the afore-mentioned process:  
\begin{equation}
\text{filter}(x_{d,i}, norm_i) = \left\{\begin{matrix}
	\text{null} &  \text{if } x_{d,i} = norm_{i}\\ 
	x_{d,i}  & \text{if } x_{d,i} \ne norm_{i}
\end{matrix}\right.
\label{eqn:filterTechnique}
\end{equation}
where the filter() function is applied to the daily input data to remove norm data, while preserving the deviated locations.
Therefore, the following Algorithm~\ref{alg:deviatedEventsGenerate} is proposed to compute for daily trajectory by iterating the filter function every day. 
\begin{algorithm}[h]
\caption{Compute Deviated \revtwo{Activity Behaviors}}
\label{alg:deviatedEventsGenerate}
\fontsize{8pt}{8pt}\selectfont
\KwData{$norm$}
\KwResult{Hybrid\_Norm $norm_{d,i}$}
\BlankLine 
\textbf{function} ComputeDeviatedEvents($X$, $norm_d$) \BlankLine \vspace{-0.11cm}
\Indp
1. Initialize $E_d$  \BlankLine \vspace{-0.11cm}
2. Compute each inputted day using filter function in Eqn.~\ref{eqn:filterTechnique}\BlankLine \vspace{-0.11cm}
\For{$i$, from $0$ to $d$}
{$E_d$ append filter($x_i \in X$, $norm_i$)}\BlankLine \vspace{-0.11cm}
\textbf{return} $E_d$					
\end{algorithm}

\begin{figure}[h]
\centering
\includegraphics[width=0.5\textwidth]{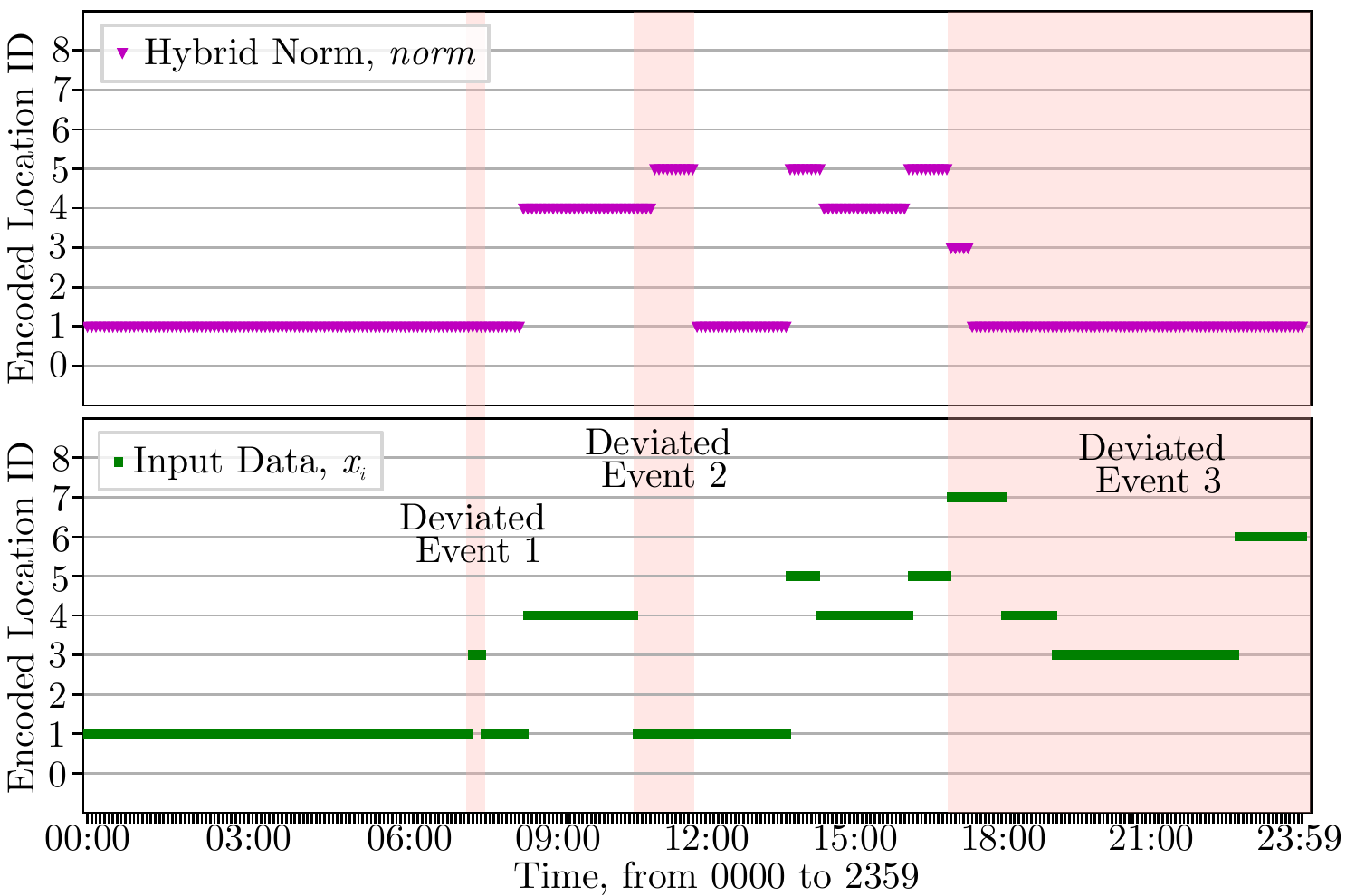} 
\vspace{-0.1cm}
\caption{Example of extracting deviated \revtwo{activity behaviors} for resident $u_{21}$ on 10 October 2019.}
\label{fig:toyExampleNorm}
\vspace{-0.36cm}
\end{figure}

Using the example stated earlier in Fig.~\ref{fig:exampleHybridExtraction}, the deviated \revtwo{activity behaviors} of resident $u_{21}$ can be computed for 10 October 2019.
The deviated \revtwo{activity behaviors} are highlighted as illustrated in Fig.~\ref{fig:toyExampleNorm}.

In the upper part of Fig.~\ref{fig:toyExampleNorm}, we show an example of obtaining a hybrid norm based on the data fusion approach by fusing group features and individual features.
After performing the filtering function, the input trajectory is highlighted in red color, which the users' trajectory is different than the computed hybrid norm.
Here, three different deviated \revtwo{activity behaviors} are detected across different locations.
The first deviated \revtwo{activity behavior} happened around 7:30am, where the resident $u_{21}$ visit level 3 public area for 15 minutes.
After that, resident $u_{21}$ return to his origin and stayed from 10:45am to 2:00pm. 
Based on the hybrid norm, the resident from cluster 4 went back to their room after 11:45pm, which is conflicted with the resident $u_{21}$'s hybrid norm.
Lastly, the last deviated \revtwo{activity behavior} occurred around 4:50pm and continued until the end of the day.
Supposedly, the resident $u_{21}$ would spend most of his/her time in origin, but instead, went to public spaces around level 3 and level 2 private area. 
This shows the process of detecting deviated \revtwo{activity behaviors} for a particular user by integrating the group norm and individual norm. 
By fusing both information, the deviated \revtwo{activity behaviors} can be computed for every resident in the nursing home and further study the types of deviated \revtwo{activity behavior}.

\begin{figure}[b]
\begin{center}
	\vspace{-0.36cm}
	\includegraphics[width=0.49\textwidth]{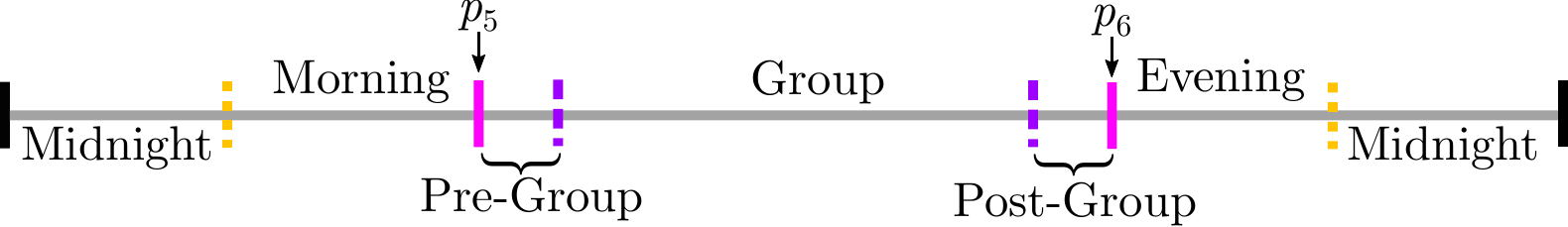} 
	\vspace{-0.5cm}
	\caption{Illustration of the finer temporal periods.}
	\label{fig:dailytimeline}
\end{center}
\end{figure}

\subsection{Analysis and Classification of Deviated \revtwo{Activity Behaviors}}
After obtaining deviated \revtwo{activity behaviors}, we proceed to identify the types of it among the nursing home's residents.
We found out some residents suffered from sleep irregularity, and some of them absent during the group activity.
Based on observation of the data, three different deviated \revtwo{activity behaviors} can be generalized, which are (1) sleep irregularity, (2) awake irregularity, and (3) private visiting.
The definition of deviated \revtwo{activity behaviors} are listed as follows:
\begin{itemize}
\item (1) sleep irregularity - The sleep irregularity denotes the resident goes to another location instead of staying back to his/her origin at the usual timing during night time.
\item (2) awake irregularity - The awake irregularity denotes the resident either leaves the room late or early compared to their routine schedule. 
\item (3) private visiting - The resident went to another staying room (staying room for level 2 or level 3) instead of the common places cluster or he/she normally will present.
\end{itemize}
Subsequently, a rules-based classification method is introduced to define the features for identifying deviated \revtwo{activity behavior}.
The rationality of using such an approach is that labels can be assigned based on the predefined rules using a statistical approach given there is no ground-truth available.
Note that each user can be assigned multiple classification labels.

\begin{table}[h]
\centering
\caption{Definition of the Transition Period}
\label{tbl:dailyTimeline}
\begin{tabular}{@{}l|l@{}}
	\toprule
	Temporal Period & Time Range of the Day \\ \midrule
	Midnight & day start $\rightarrow$ ($\frac{p_5}{2}$) \\
	Morning & $\frac{p_5}{2}\rightarrow p_5$\\
	Pre-group & $p_5\rightarrow (p_5 + 60$mins$)$\\
	Group & $(p_5 + 60$mins$)\rightarrow (p_5 - 60$mins$)$\\
	Post-group & $(p_6 - 60$mins$)\rightarrow p_6$\\
	Evening &  $p_6 \rightarrow (\frac{t=288 - p_6}{2}$)\\
	Midnight &  $(\frac{p_6}{2}) \rightarrow$ day end \\ \bottomrule
\end{tabular}
\end{table}

\begin{figure}[b!]
\begin{center}
	\vspace{-0.36cm}
	\includegraphics[width=0.48\textwidth]{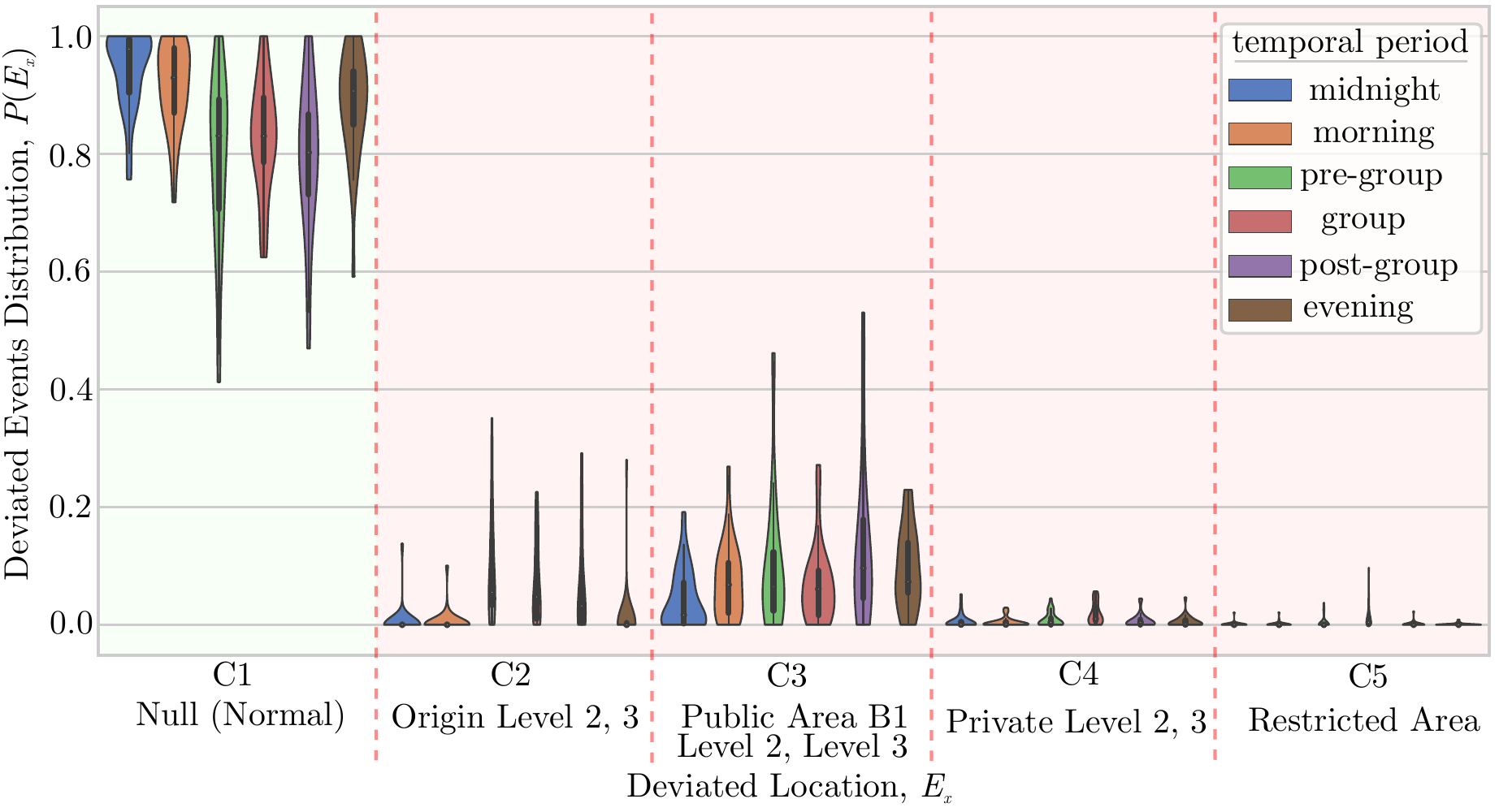} 
	\vspace{-0.6cm}
	\caption{Violin plot for the deviated location w.r.t. different temporal periods. Note that c1 denotes probability of residents being normal while c2 to c5 represent the deviated locations based on their functionalities.}
	\label{fig:distribution}
\end{center}
\vspace{-0.36cm}
\end{figure}

To generate features for the classification, we further break down the time of the day into seven different finer temporal periods as shown in Fig.~\ref{fig:dailytimeline}.
By breaking down into different periods, the time before and after group activity can be investigated to generate features for studying deviated \revtwo{activity behavior}.
Note that $p_5$ and $p_6$ shown here are based on the average time start and time end throughout the data collection period for each user.
Using Fig.~\ref{fig:dailytimeline} as a reference, the finer temporal periods are introduced as shown in Table~\ref{tbl:dailyTimeline}.

Next, the probabilities of deviated locations' occurrence are calculated based on the daily filtered deviated location for each finer temporal period.
This can be used to define rules using the probabilistic calculation based on the deviated \revtwo{activity behavior} statistic.
The location probability, $P(E_{i,x})$ can be computed using following Eqn.~\ref{eqn:probDeviatedEvents}
\begin{equation}
P(E_{i,x}) = \frac{q(E_{i,x})}{c} 
\label{eqn:probDeviatedEvents}
\end{equation}
where $i$ denotes the timeslot of the day and $x$ represents the location resident went when a deviated behavior occurred.
The $q()$ counts the occurrence number of deviated \revtwo{activity behavior} $E_{i,x}$ and $c$ denotes the total number of the valid time slot.
By combining the probability of deviated locations of every user, we can study the potential timeslot for each user, where the deviated \revtwo{activity behaviors} occurred most.
To simplify the locations of where deviated \revtwo{activity behaviors} occurred, encoded locations with similar functionalities are combined and study the probability for each timeslot.
The combined encoded list are listed as follows: (c1) null, (c2) origin, (c3) public area, (c4) private area, and (c5) restricted location. 

\begin{figure}[t!]
\begin{center}
	%		\includegraphics[width=0.49\textwidth]{FIGURES/abnormalTemporalDistribution} 
	%		\vspace{-0.6cm}
	%		\caption{\rev{Violin plot for the deviated location w.r.t. different temporal periods.}}
	%		\label{fig:distribution}
	
	\includegraphics[width=0.49\textwidth]{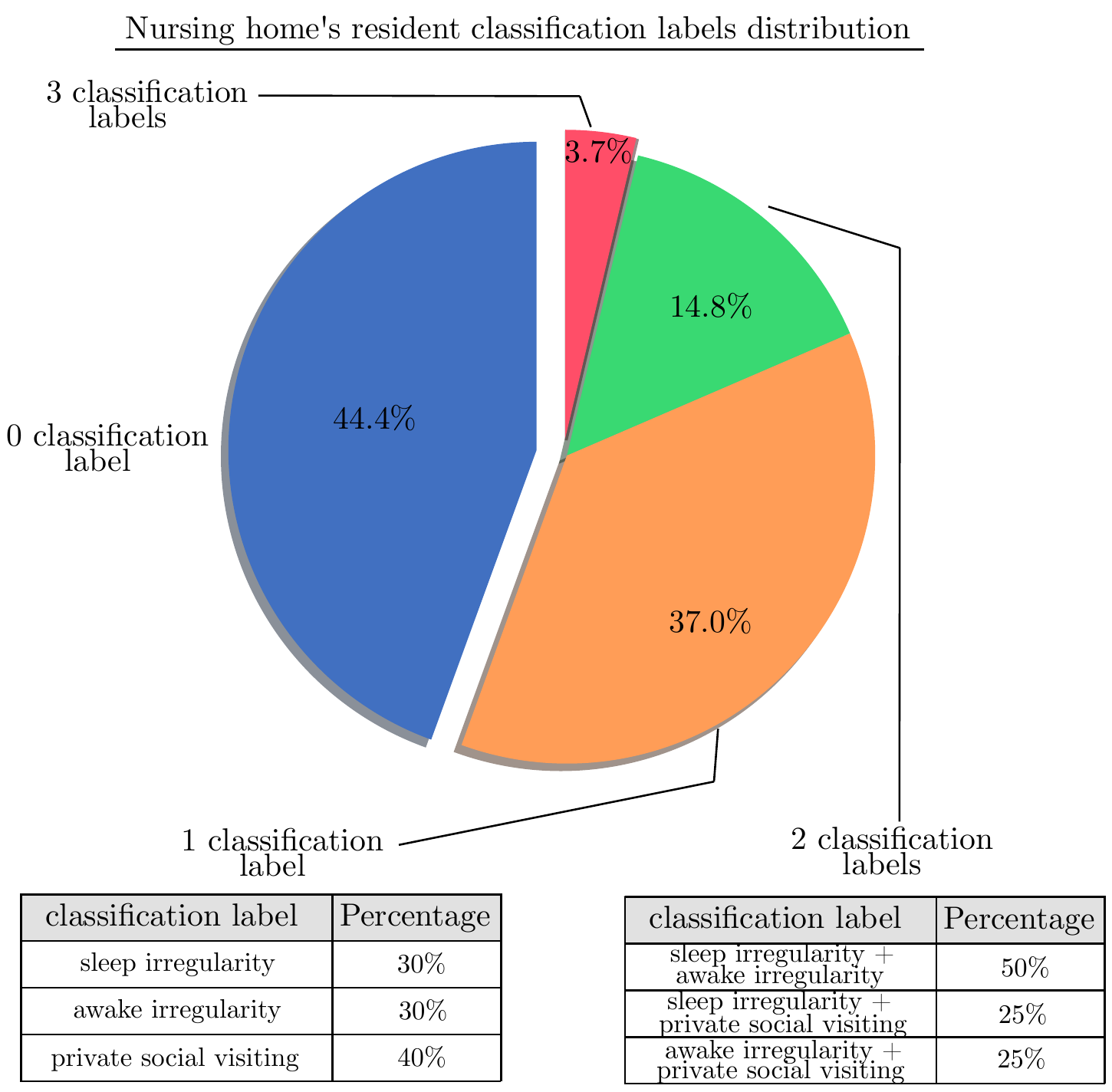} 
	\caption{Distribution of residents based on the number of classification labels.}
	\label{fig:rules_general_result} 
	\vspace{-0.6cm}
\end{center}
\end{figure}

\begin{figure}[h!]
\begin{center}
	\begin{tabular}{@{}c@{}}
		(a) Example of awake irregularity detected for resident $u_{35}$\\
		\includegraphics[width=0.49\textwidth]{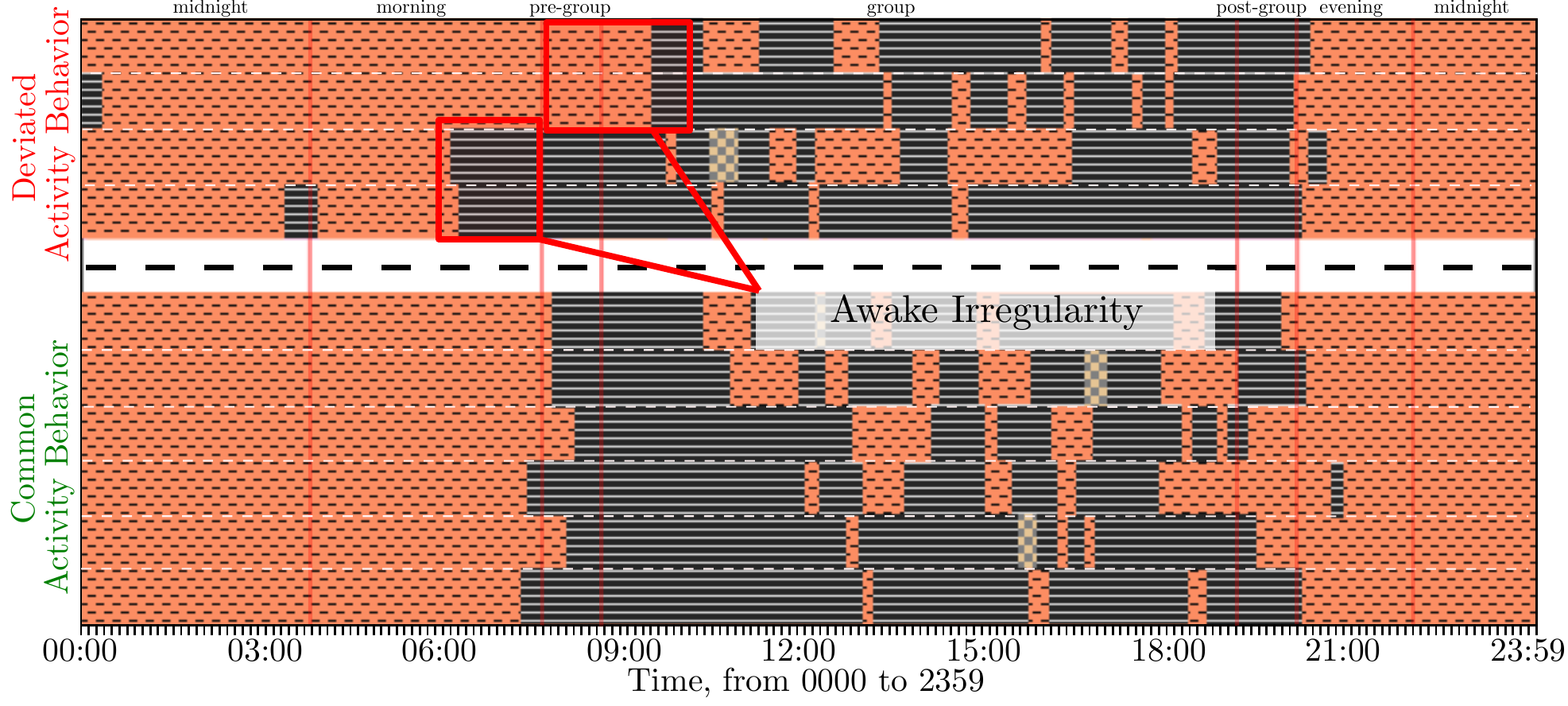} \\
		(b) Example of sleep irregularity detected for resident $u_{5}$ \\
		\includegraphics[width=0.49\textwidth]{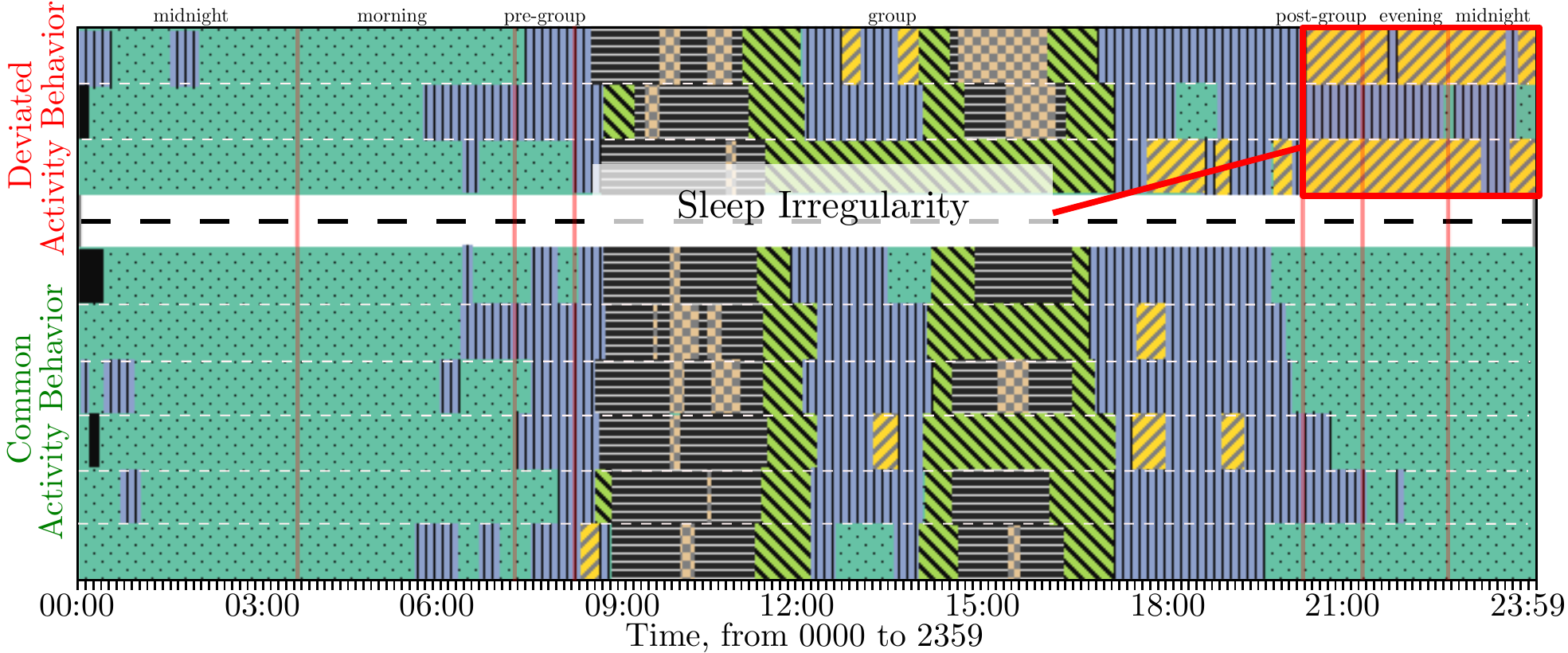} \\
		(c) Example of private visiting detected for resident $u_{21}$\\
		\includegraphics[width=0.49\textwidth]{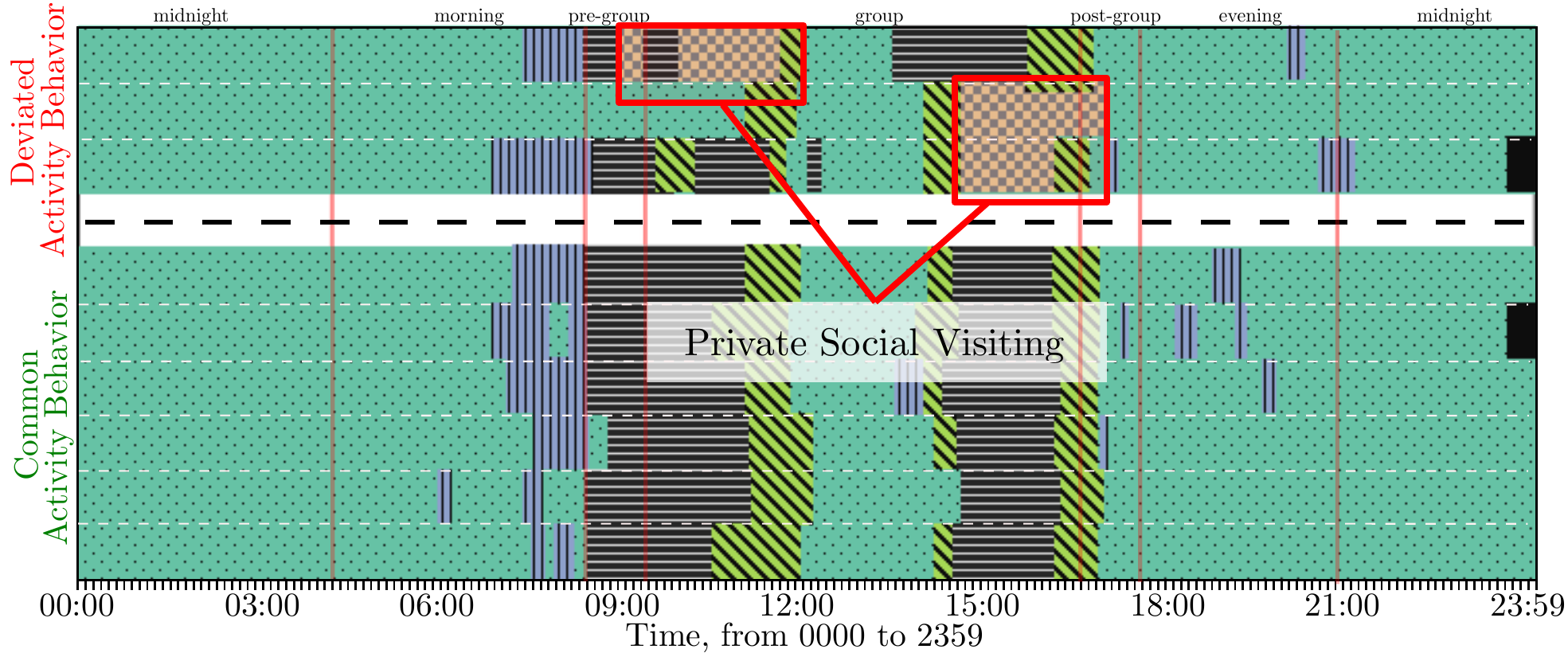} \\
		\includegraphics[width=0.4\textwidth]{FIGURES/Spect_Legends}
	\end{tabular}
	\caption{Example of the deviated \revtwo{activity behavior} study. Note that the y-axis consists of sample data extracted user trajectory on selected day. The common \revtwo{activity behavior} denotes trajectory data without any deviated \revtwo{activity behavior}, while the deviated \revtwo{activity behavior} category represents the opposite.}
	\label{fig:deviatedBehaviorFocus} 
\end{center}
\vspace{-0.36cm}
\end{figure}

After computing the probabilities of deviated locations for the nursing home's residents, we illustrate the distribution of deviated locations using a violin plot as shown in Fig.~\ref{fig:distribution}.
The category c1 represents the probability of residents not involving in deviated \revtwo{activity behavior} at a different time of the day, where category c2 to c5 represents the location for a deviated \revtwo{activity behavior} to be detected.
The probability of deviated \revtwo{activity behavior} occur at other locations c2 to c5 is small which the value ranges from $0.0$ to $0.4$.
One can observe a lot of deviated \revtwo{activity behaviors} that happened mostly in the public area (c3) and follow by origin (c2).
That being said, to further understand those involved in sleep irregularity and social visiting events, attention is given to the residents who yield a higher probability of deviated \revtwo{activity behaviors}.
Specifically, we are interested in the distribution of residents, whose probability is higher than the median for the encoded location c2 to c5.
Those residents have higher chances of showing deviated \revtwo{activity behaviors} compared to the median of the distribution. 
Using the characteristic of a violin plot, the classification threshold is defined based on the upper adjustment value (UAV) and lower adjustment value (LAV), which are computed through the first or third quantile $\pm 1.5\times$ interquartile range.
Based on the specific timing's distribution, the UAV and LAV values are extracted as shown in Table~\ref{tbl:thresholdValue} to define the threshold values for the classification rules.
\begin{table}[h]
\caption{UAV and LAV Extracted for Classification Rules}
\centering
\begin{tabular}{c|c|c|c}
	\hline			
	\multirow{2}{*}{Deviated Location} & c1 & c2 & c4 \\ \cline{2-4} 
	& LAV & UAV & UAV \\ \hline
	midnight & 0.8762 &-&-\\
	Morning & 0.7323 &-& 0.0118 \\
	pre-group &-&0.2717& 0.0312 \\
	group &-&-& 0.0521 \\
	post-group &-&-& 0.0223 \\
	evening & 0.7522 &-& 0.0174 \\ \hline
\end{tabular}
\label{tbl:thresholdValue}
\end{table}

Based on the input from Table~\ref{tbl:thresholdValue}, we define the classification rules as presented in Table~\ref{tbl_classificationRules}.
\begin{table}[]
\caption{Deviated \revtwo{Activity Behavior} Classification Rules}
\centering
\begin{tabular}{@{}l|c@{}}
	\toprule
	Deviated \revtwo{Activity Behavior} & Classification Rules \\ \midrule
	Sleep Irregularity &  c1 midnight $<$ LAV \text{or} c1 Evening $<$ LAV\\  \midrule
	Awake Irregularity &  c1 morning $<$ LAV \text{or} c2 pre-group $<$ LAV\\  \midrule
	Private Visiting & \begin{tabular}[c]{@{}c@{}} c4 morning $>$ UAV \text{or} \\
		c4 pre-group $>$ UAV \text{or} \\
		c4 group $>$ UAV \text{or} \\
		c4 post-group $>$ UAV \\
		\text{or} c4 evening $>$ UAV
	\end{tabular} \\ \bottomrule
\end{tabular}
\label{tbl_classificationRules}
\end{table}

After defining the classification rules, we classify every nursing home's residents based on their probability for the finer hybrid temporal period.
A general statistic for the classified labels after performing the rules-based classification is presented in Fig.~\ref{fig:rules_general_result}.
Based on the result, 44.4\% of residents do not have a classification label, which indicates the majority of the residents do not have deviated \revtwo{activity behavior}.
The remaining 37\% of the residents are associated with one classification label only, of which 40\% of them are having private social visiting irregularity.
On the other hand, half of the residents with two classification labels mainly consist of sleep and awake irregularity.

Based on the classification labels generated, we examine three residents as case studies to study normal daily \revtwo{activity behavior} and deviated \revtwo{activity behavior}.
Three different deviated \revtwo{activity behaviors} are illustrated in Fig.~\ref{fig:deviatedBehaviorFocus}.
From Fig.~\ref{fig:deviatedBehaviorFocus}(a), the resident $u_{35}$ is not waking up based on his/her daily schedule. 
Based on observation, there are two types of awake irregularity, which are wake up earlier or wake up way later than the normal wake up time (around 7:30 am). 
Next, we investigate the sleep irregularity of resident $u_{5}$ as depicted in Fig.~\ref{fig:deviatedBehaviorFocus}(b).
The resident $u_5$ normally goes back to the origin after 9:00pm, but in the deviated \revtwo{activity behavior}'s extracted days, he/she remained at the public area until the end of the day.
These \revtwo{activity behaviors} normally do not occur in his/her normal routine, and this is a good representation of the sleep irregularity.
Subsequently in Fig.~\ref{fig:deviatedBehaviorFocus}(c), one can discover that the resident $u_{21}$ absent during the group activity for few days, where he/she went to a private area instead.
This phenomenon indicates the resident $u_{21}$ either is having social interaction with other residents in their private space or avoiding group activity. 
This could be potentially a deviated \revtwo{activity behavior}.

By utilizing the proposed hybrid deviated \revtwo{activity behavior} classification, we managed to obtain types of residents' deviated \revtwo{activity behaviors} in a nursing home. 
This provides insight into the nursing home's management regarding the residents' deviated \revtwo{activity behavior} and attention can be provided to the individual with needs.

%%%%%%%%%%%%%%%%%%%%%%%%%%%%%%%%%%%%%%%%%%%%%%%%%%%%%%%%%%%%%%%%%%%%%%%%%%%%%%%%%%%%%%%%%%%%%%%%%%%%%%%%%
\section{Conclusion}
\label{sec:conclusion}

In this paper, we present a location-based deviated \revtwo{activity behavior} detection system for the Salvation Army, Peacehaven Nursing Home, Singapore.
The 50 residents can be segmented into different groups based on their \revtwo{activity behavior}, which contributes to formulating group norms.
By combining group and individual norms, we can generate a hybrid norm to identify the common behavior for each resident.
By understanding residents' normal \revtwo{activity behavior}, the deviated \revtwo{activity behavior} can be differentiated and extracted from normal \revtwo{activity behavior} to study it. 
Based on the types of deviated \revtwo{activity behavior}, three categories of deviated \revtwo{activity behavior} are proposed, which are (1) sleep irregularity, (2) awake irregularity, and (3) private visiting. 
Next, three users' normal and deviated \revtwo{activity behavior} are studied after performing rules-based classification.

For future works, we plan to incorporate more data sources to generate more accurate deviated \revtwo{activity behavior} detection.
Also, a real-time deviated \revtwo{activity behavior} system is part of the future research direction as the dynamic formation of group activity is yet another issue to address.
\revtwo{Another aspect that can be improved is getting ground-truth for the data collected to further enhancing deviated activity behavior detection module.} 
\revtwo{Moreover, a more throughout study on the compatibility with different countries' data protection rules would be part of the research interest to reach out to more nursing homes.}

%%%%%%%%%%%%%%%%%

%%%%%%%%%%%%%%%%%%%%%%%%%%%%%%%%%%%%%%%%%%%%%%%%%%%%%%%%%%%%%%%%%%%%%%%%%%%%%%%%%%%%%%%%%%%%%%%%%%%%%%%%%%%
%% use section* for acknowledgment
%\section*{Acknowledgment}
%%%%%%%%%%%%%%%%%%%%%%%%%%%%%%%%%%%%%%%%%%%%%%%%%%%%%%%%%%%%%%%%%%%%%%%%%%%%%%%%%%%%%%%%%%%%%%%%%%%%%%%%%%%
%This work is supported by Singapore Ministry of National Development (MND) Sustainable Urban Living Program, under the grant no. SUL2013-5, ``Liveable Places: A Building Environment Modeling Approach for Dynamic Place Making” project, and especially appreciate the useful discussion and help from the collaborators from MND, HDB and URA.

\bibliographystyle{IEEEtran}
\newcommand{\BIBdecl}{\setlength{\itemsep}{0.25 em}}
\bibliography{bare_journal}

\end{document}